\documentclass[preprint]{aastex}
\usepackage{amsmath}
\usepackage{natbib}
\usepackage{epstopdf}
\usepackage{multirow}

\newcommand{\mppc}[1]{\:M_\odot {\rm pc}{}^{-#1}}
\def\del{\partial}

\begin{document}
\title{Interstellar Gas and a Dark Disk}
\author{Eric David Kramer and Lisa Randall}
\affil{Department of Physics, Harvard University, Cambridge, MA, 02138}

\begin{abstract}
We introduce a potentially powerful method for constraining or discovering a thin dark matter disk in the Milky Way. The method relies on the relationship between the midplane densities and scale heights of interstellar gas being determined by the gravitational potential, which is sensitive to the presence of a dark disk. We show how to use the interstellar gas parameters to set a bound on a dark disk and discuss the constraints suggested by the current data. However, current measurements for these parameters are discordant, with the uncertainty in the constraint being dominated by the molecular hydrogen midplane density measurement, as well as by the atomic hydrogen velocity dispersion measurement. Magnetic fields and cosmic ray pressure, which are expected to play a role, are uncertain as well. The current models and data are inadequate to determine the disk's existence, but, taken at face value, may favor its existence depending on the gas parameters used.
\end{abstract}

\section{Introduction}

 Fan, Katz, Randall, and Reece in 2013 proposed the existence of thin disks of dark matter in {spiral galaxies including the Milky Way}, in a model termed Double Disk Dark Matter (DDDM). In this model, a small fraction of the dark matter is interacting and dissipative, so that this sector of dark matter would cool and form a thin disk. More recently \citet{dino} showed that a dark matter disk of surface density $\sim 10 \mppc{2}$ and scale height $\sim$10 pc could possibly explain the periodicity of comet impacts on earth.  It is of interest to know what values of dark disk surface density and scale height are allowed by the current data, and whether these particular values are allowed.

Since the original studies by \citet{oort1,oort2}, the question of disk dark matter has been a subject of controversy. Over the years, several authors have suggested a dark disk to explain various phenomena. \citet{kalb07} proposed a thick dark disk as a way to explain the flaring of the interstellar gas layer. It has also been argued that a thick dark disk is formed naturally in a $\Lambda$CDM cosmology as a consequence of sattelite mergers \citep{dd1}. Besides these, there are also models arguing for a thin dark disk. \citet{dddm} put forward a model for dark matter where a small fraction of the total dark matter could be self-interacting and dissipative, necessarily forming a thin dark disk. In \citet{paper1} we investigated the constraints on such a disk from stellar kinematics. In this paper we investigate the contraint imposed by demanding consistency between measurements of midplane densities and surface densities interstellar gas.

We assume a Bahcall-type model for the vertical distributions of stars and gas \citep{bahc84c,bahc84b,bahc84a} as in \citet{paper1}, with various visible mass components, as well as a dark disk. We investigate the visible components in detail given more recent measurements of both the surface and midplane densities. A dark disk affects the relationship between the two as argued in \citeauthor{paper1} and as we review below. Although current measurements are insufficiently reliable to place strong constraints on or identify a disk, we expect this method will be useful in the future when better measurements are achieved.

\section{Poisson-Jeans Theory}

As explained in detail in \citet{paper1}, for an axisymmetric self-gravitating system, the vertical Jeans equation near the $z=0$ plane reads
\begin{equation}
\label{eq:poissonjeans0}
\frac{\del}{\del z}(\rho_i \sigma_i^2) + \rho_i \frac{\del \Phi}{\del z} = 0.
\end{equation}
For an isothermal population ($\sigma_i(z) = {\rm constant}$), the solution reduces to
\begin{equation}
\label{eq:ideal}
\rho_i(z)=\rho_i(0)\,e^{-{\Phi(R,z)}/{\sigma_i^2}}.
\end{equation}
Combining this with the Poisson equation gives the Poisson-Jeans equation for the potential $\Phi$
\begin{equation}
\label{eq:poissonjeans1}
\frac{\del^2\Phi}{\del z^2} = 4\pi G \sum_i \rho_i(0) e^{-\Phi/\sigma_i^2},
\end{equation}
which can also be cast in integral form (assuming $z$-reflection symmetry)
\begin{equation}
\label{eq:poissonjeans2}
\frac{\rho_i(z)}{\rho_i(0)} = \exp\left(-\frac{4\pi G}{\sigma_i^2}\sum_k \int_0^{z}\!dz^{\prime}\,\int_0^{z^{\prime}}\!dz^{\prime\prime}\, \rho(z^{\prime\prime})\right).
\end{equation}
This is the form used in our Poisson-Jeans solver.

\subsection{A toy model}
\label{sec:toy}
\begin{figure}[h]
\plotone{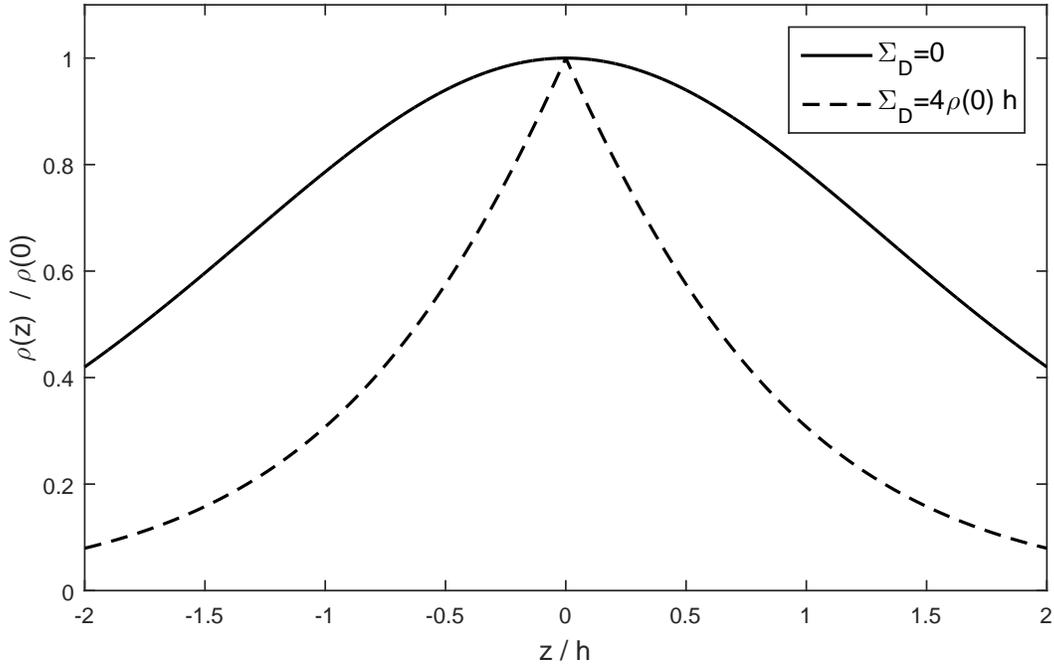}
\centering{}\caption{{  A plot of the exact solutions without and with a dark disk of $Q=1$. The density is `pinched' by the disk, in accordance with Equation \ref{eq:exact}.}}
\label{fig:toy}
\end{figure}

In \citet{paper1}, we showed that the exact solution to the Poisson-Jeans equation for a thick component (in this case, the interstellar gas which is thick compared to the dark disk) with midplane density $\rho_0$ and vertical dispersion $\sigma$ interacting with an {infinitely thin (delta-function profile)} dark disk was 
\begin{equation}
\label{eq:exact}
\rho(z)=\rho_{0} (1+Q^2)\; {\rm sech}^2\!\left( \frac{\sqrt{1+Q^2}}{2h}\left(|z| + z_0\right)\right)
\end{equation}
where 
\begin{equation}Q\equiv \Sigma_D/4\rho_{0}h,\end{equation}
\begin{equation}h\equiv \frac{\sigma}{\sqrt{8\pi G\rho_0}},\label{eq:disp}\end{equation}
and
\begin{equation}z_0 \equiv \frac{2h}{\sqrt{1+Q^2}} {\rm arctanh}\left(\frac{Q}{\sqrt{1+Q^2}}\right).\end{equation}
Thus, the effect of the dark disk is to `pinch' the density distribution of the other components, as we can see in Figure \ref{fig:toy}. Thus, although the scale height of the gas disk is proportional to its velocity dispersion according to Equation
 \ref{eq:disp}, a dark disk will reduce the gas disk's thickness relative to this value, and for a fixed midplane density $\rho_i(0)$, it implies that their surface densities $\Sigma_i$ are less than what it would be without the dark disk or any other mass component. In this approximation, the gas distribution will have a cusp at the origin, but in general, the dark disk will have a finite thickness and the solution will be smooth near $z=0$.

Integrating (\ref{eq:exact}) gives the surface density of the visible component as
\begin{equation}
\begin{split}
\Sigma_{\rm vis}(\Sigma_D)=\sqrt{\Sigma_{\rm vis}(0)^2+\Sigma_D^2}-\Sigma_D
\end{split}
\end{equation}
where $\Sigma_{\rm vis}(0)\equiv 4\rho_{0}h$ is what the surface density would have been without the dark disk. This expression is monotonically decreasing with $\Sigma_D$.

Another way of explaining this is that the dark disk `pinches' the visible matter disk, reducing its thickness $H_{\rm vis}$. Since the surface density of the visible disk scales roughly as $\Sigma_{\rm vis}\sim \rho_{\rm vis}\, H_{\rm vis}$, the effect of the dark disk is to reduce the total surface density for a given midplane density $\rho_{\rm vis}$.

\section{Analysis}
\label{sec:analysis}

Here we explain how we compare the surface densities of the various gas components estimated in the next section to those predicted by their midplane densities and velocity dispersions in order to place self-consistency constraints on the mass model. Section \ref{sec:toy} explains how the presence of the dark disk decreases the surface density of each component if the midplane density is held fixed (as it is in a Poisson-Jeans solver). Thus, given fixed midplane densities, we can assign a probability to a model with any dark disk surface density $\Sigma_D$ and scale height $h_D$ based on how well the predicted surface densities $\Sigma_i$ determined from the Poisson-Jeans solver match the observed values.

Starting with the midplane densities and dispersions of Section \ref{sec:massmodel}, we solved the Poisson-Jeans equation for $\Sigma_D$ values between 0 and 24 $\mppc{2}$. Each time the Poisson-Jeans equation was solved, the density distributions were integrated to give the total surface densities of $\rm H_2$ and HI. Each model was then assigned a probability, according to the chi-squared distribution with 2 degrees of freedom, based on the deviation of this surface density from the measured values. We did this using different central values and uncertainties for the midplane densities. Thus, for example, using $n_{{}_{\rm H_2}}=0.19\,{\rm cm^{-3}}$, we would take $\rho_{{}_{\rm H_2+He}}(0)= 1.42\times m_{{}_{\rm H_2}} \times 0.19 \,{\rm cm^{-3}}=0.013 \mppc{3}$. If for a model with this value of $\rho_{\rm H_2}(0)$ and with a certain dark disk surface density value of $\Sigma_D$ we find an H${}_2$ surface density $\Sigma_{\rm H_2}=1.0 \mppc{2}$, then according to Section \ref{sec:h2}, we should assign this model a chi-squared value $\chi^2_{\rm H_2}=(1.0-1.3)^2/\Delta_{\Sigma_{\rm H2}}^2$. We would then assign a probability to this model according to the Gaussian cumulative distribution, $p_{\rm H_2}=\int\!d\chi\, \exp(-\chi^2/2)/\sqrt{2\pi}$, where the limits of integration are from $-\infty$ to $-\chi_{\rm H_2}$ and $\chi_{\rm H_2}$ to $\infty$. We would similarly compute probabilities $p_{\rm HI}$, $p_{\rm HII}$, and a combined probability $p=p_{\rm H_2}\times p_{\rm HI}\times p_{\rm HII}$. We note here that this is not the absolute probability of the model given the data; rather, it is the probability of the data given the model. We define a model for which the data is less probable than 5\% to be excluded.

An important question is what to use for $\Delta_{\Sigma}^2$. There are two uncertainties here. Namely, 1) the uncertainty in the surface density measurements, $\Delta_{\hat\Sigma_i}$, and 2) the uncertainty in our output values of $\Sigma(\hat\rho_i,\Sigma_D)$, resulting from the uncertainty in the input midplane density measurements $\hat\rho_i$. Formally, this is $|\partial \Sigma_i /\partial \rho_i|\Delta_{\hat\rho_i}$. Assuming Gaussian distributions for the measurements $\hat\Sigma_i$ and $\hat\rho_i$, and a uniform prior for $\Sigma_D$, one can show that 
\begin{equation}
p(\hat\Sigma_i,\hat\rho_i|\Sigma_D)\sim \int\!\! d \rho_i \exp\left(-\frac{\left(\hat\Sigma_i-\Sigma(\rho_i,\Sigma_D)\right)^2}{2\Delta_{\hat\Sigma_i}^2}\right)\exp\left(-\frac{\left(\rho_i-\hat\rho_i\right)^2}{2\Delta_{\rho_i}^2}\right)
\end{equation}
where $\rho_i$ are the true midplane densities, and that, expanding $\Sigma(\rho_i,\Sigma_D)$ to first order in $\rho_i$, this integrates to give an approximately Gaussian distribution for $\hat\Sigma_i-\Sigma(\hat{\rho_i},\Sigma_D)$, with width
\begin{equation}
\label{eq:Delta}
\Delta_{\Sigma_i}\simeq \sqrt{\Delta_{\hat\Sigma_i}^2+\left(\frac{\partial \Sigma_i}{\partial \rho_i}\right)^2\!\Delta_{\hat\rho_i}^2}.
\end{equation}
We computed $|\partial \Sigma_i/\partial \rho_i|$ by sampling values of $\rho_i$ and computing the output values $\Sigma_i$. The effect of the uncertainties in the vertical dispersions of the different components was also included in $\Delta_{\Sigma_i}$ in the same way as those in $\hat\rho_i$. The formula for $\Delta_{\Sigma_i}$ that we adopt (Equation \ref{eq:Delta}) therefore contains an extra term under the square root to include this uncertainty:
\begin{equation}
\label{eq:Delta2}
\Delta_{\Sigma_i}\simeq \sqrt{\Delta_{\hat\Sigma_i}^2+\left(\frac{\partial \Sigma_i}{\partial \rho_i}\right)^2\!\!\Delta_{\hat\rho_i}^2+\left(\frac{\partial \Sigma_i}{\partial \sigma_{i,{\rm eff}}}\right)^2\!\!\Delta_{\hat\sigma_{i,{\rm eff}}}^2}.
\end{equation}

\section{Gas parameters}
\label{sec:massmodel}

The purpose of this section is to determine accurate values for $\rho_i(0)$, $\sigma_i$, and $\Sigma_i$ (midpalne density, velocity dispersion, and surface density) for the different components of interstellar gas based on existing measurements. Using Equation \ref{eq:ideal}, these can then be compared for a given dark disk model in order to check for self-consistency.

We now discuss in detail the various measurements of the gas parameters and the uncertainties in each. Our starting point is the Bahcall model used by \citet[Table 2]{hf2006}. These values are updated from the ones used in \citet{hf2000}. Values for the stellar components were updated using the values of \citet{mckee}, and are shown in rows 5-15 in Table 2 of \citet{paper1}.

In these models, the gas and stars are both separated into approximately isothermal components as in \citet{bahc84b}, so that each component $i$ is characterized by a midplane density $\rho_{i0}$ and a vertical dispersion $\sigma_i$. Using only these values for all of the components, we can solve the Poisson-Jeans equation (\ref{eq:poissonjeans2}) for the system. A major difference between our model and that of \citet{hf2006} is that their gas midplane densities were fixed by the values needed to give the correct surface densities in accordance with the Poisson-Jeans equation. We, on the other hand, use measured values of the midplane densities as we explain in this section.

We explain the various literature values that were included in the determination of the gas parameters. We also compare these to the recent values of \citet{mckee}. In Section \ref{sec:results}, the analysis is conducted separately for the values we determine by combining the results in the literature and the values obtained solely from the recent paper by \citet{mckee}.

%

\subsection{Molecular hydrogen}
\label{sec:h2}

We now explain the various measurements of the molecular hydrogen volume density and surface densities and how they are corrected. As molecular hydrogen cannot be observed directly, it must be inferred from the amount of CO present, derived from the intensity of the $J=1-0$ transition photons. These are related by the so-called $X$-factor, defined by 
\begin{equation}
\label{eq:X}
N_{\rm H_2}\equiv X W_{\rm CO}
\end{equation}
where $N_{\rm H_2}=N_{\rm l.o.s.}$ is the line-of-sight column density of $\rm H_2$ molecules and $W_{\rm CO}$ is the total, velocity-integrated CO intensity along the line of sight \citep{draine}. Column densities perpendicular to the galactic plane can then be obtained by simple trigonometry: 
\begin{equation}
N_\perp = N_{\rm l.o.s.} \sin b
\end{equation}
and volume densities can be obtained by dividing the intensity density in velocity space $dW_{\rm CO}/dv$ by the rotation curve gradient $dv/dR$, or by estimating the distance along the line of sight using other means. The volume and surface densities can also both be found by fitting an assumed distribution to measurements of the gas' vertical scale height $\Delta z$. Surface densities can then be given, for example, by
\begin{equation}
\Sigma_{\rm H_2}=m_{{}_{\rm H_2}}\,N_{\rm \perp, H_2}=m_{{}_{\rm H_2}} X W_{\rm CO} \sin b.
\end{equation}
On the other hand, a certain reference may not be measuring surface density directly. Instead, they may be measuring the emissivity,
\begin{equation}
J({\bf r})\equiv\frac{dW_{\rm CO}}{dr}
\end{equation}
from which, according to Equation \ref{eq:X}, we can obtain the volume density as
\begin{equation}
n({\bf r})=X\,J({\bf r}).
\end{equation}
If the authors also measured the vertical ($z-$direction) distribution of the molecular hydrogen, then the surface mass density can be obtained according to
\begin{equation}
\Sigma_{\rm H_2}=m_{{}_{\rm H_2}}\int n_{{}_{\rm H_2}}(z)\,dz.
\end{equation}
For example, the full width at half maximum (FWHM) of the molecular hydrogen distribution gives the surface density as
\begin{equation}
\label{eq:CnF}
\Sigma_{\rm H_2}=m_{{}_{\rm H_2}} C_{\rm shape}\, n_{{}_{\rm H_2}}(0)\times {\rm FWHM}
\end{equation}
where $C_{\rm shape}$ is given by 1.06, 1.13, or 1.44 for a Gaussian, $\rm sech^2$, or exponential profile respectively. For our calculations, we used
\begin{equation}
C_{\rm shape}=1.10
\end{equation}
as a reasonable estimate for the shape of the distribution.

In the literature, mass values are often quoted including the associated helium, metals, and other gaseous components such as CO, etc. The amount of helium accompanying the hydrogen is typically assumed in the range 36-40\% of the hydrogen alone by mass \citep{ismkh,bronfman}. Including other gas components increases this number to about 42\% \citep{ferriere}. Thus, the total mass of any component of the ISM should be about 1.42 times the mass of its hydrogen. These will be distinguished by using, e.g. $\Sigma_{\rm H_2}$, $\Sigma_{\rm H_2 + He}$ to refer to the bare values and and the values including their associated helium respectively. Thus,
\begin{equation}
\Sigma_{\rm H_2+He}=1.42\,\Sigma_{\rm H_2}.
\end{equation}•
Note that \citet[p.662]{bm} did not include helium in the total ISM mass. Also \citet{readrev} did not distinguish between HI results including and not including helium. 

\begin{figure}[h]
\centering{}\caption{{ Molecular hydrogen midplane densities and surface densities determined by various authors between 1984 and 2006.}}
\plotone{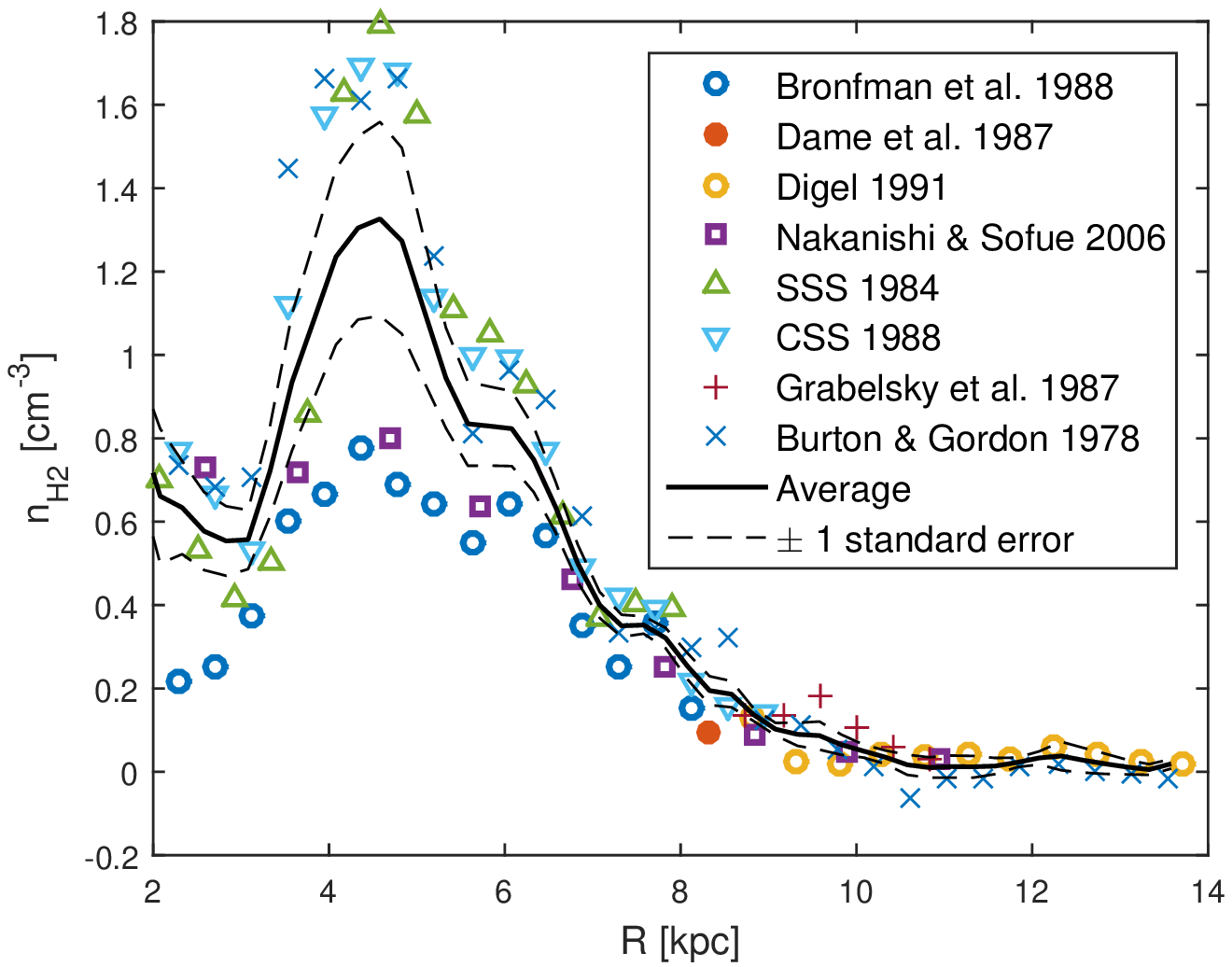}\\
\plotone{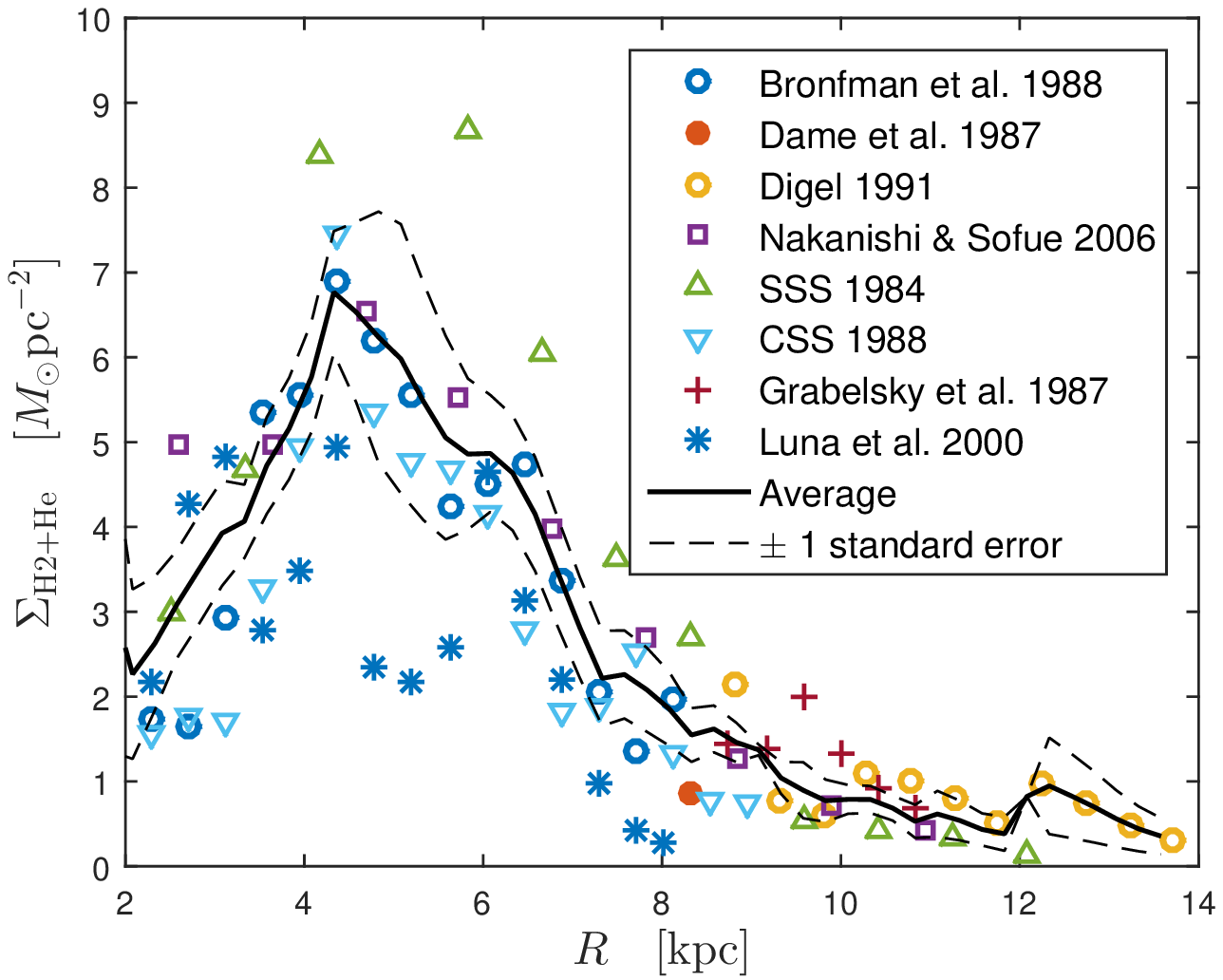}
\label{fig:nh2}
\end{figure}

We now explain how we obtain midplane volume densities $n_{{}_{\rm H_2}}(z=0)$ and surface densities $\Sigma_{\rm H_2 + He}$ from the various references in the literature. \citet{bronfman} measured the molecular hydrogen over different radii within the solar circle. Their data are shown as one of the data sets in Figure \ref{fig:nh2}. Averaging the values from the Northern and Souther Galactic plane in Table 4 of the latter, we find, for the measurements closest to the Sun, $\Sigma_{\rm H_2}=2.2\mppc{2}$ and $n_{\rm H_2}=0.2\,{\rm cm^{-3}}$. Since surface densities depend only on the total integrated intensity along the line of sight, they are independent of the value of $R_\odot$, the Sun's radial position from the center of the Galaxy. On the other hand, it follows from this that old values for volume densities (which scale as $R_\odot^{-1}$) must be rescaled by $R_{\odot,{\rm old}}/R_{\odot,{\rm new}}$ \citep[p.31]{ismss}. Since \citeauthor{bronfman} used the old value $R_\odot=10$ kpc, this value needs to be rescaled by $(0.833)^{-1}$ to take into account the new value of $R_\odot= 8.33\pm0.35$ kpc \citep{gillesen}. They also used an $X$-factor of $X=2.8\times 10^{20}\,{\rm cm^{-2}\, (K^{-1}\, km\, s^{-1})^{-1}}$. We correct this using a more recent value of $X=1.8\pm 0.3\times 10^{20}\,{\rm cm^{-2}\, (K^{-1}\, km\, s^{-1})^{-1}}$, obtained by \citet*{dht}. The most recent value of $X$, obtained by Okumura \& Kamae (\citeyear{okumura09}), is $X=1.76\pm0.04\times 10^{20}\,{\rm cm^{-2}\, (K^{-1}\, km\, s^{-1})^{-1}}$, although the value of Dame et al. that we use is still cited by \citet{draine} as the most reliable. These corrections give $n_{\rm H_2}=0.15\mppc{3}$ and $\Sigma_{\rm H_2}=1.4\mppc{2}$. Including helium gives $\Sigma_{\rm H_2+He}=2.0\mppc{2}$.

On the other hand, \citet*{css}, found the local CO emissivity $J=dW_{\rm CO}/dr$ in the first galactic quadrant for radii through $R_\odot$. For $R<R_\odot$ and $R>R_\odot$ respectively, they found these to be $J = {\rm 3.1}$ and ${\rm 2.3\; K\; km\; s^{-1}\; kpc^{-1}}$, which, using $X=1.8\times 10^{20}$ ${\rm cm^{-2}\, (K^{-1}\, km\, s^{-1})^{-1}}$, and rescaling for $R_\odot$ by $(0.833)^{-1}$, gives interpolated density $n_{\rm H_2}(R_\odot)=0.19\, {\rm cm^{-3}}$. Using their {\rm FWHM} measurements for $\rm H_2$, we can convert their measurements to surface density values according to Equation \ref{eq:CnF}. As before, the surface density values are independent of $R_\odot$. We have, interpolating to $R_\odot$, $\Sigma_{\rm H_2+He}= 1.1 \mppc{2}$. The rescaled data are shown in Figure \ref{fig:nh2}. 

Another measurement is provided by \citet{bg1978}, who had already measured Galactic CO emissivity $J=dW_{\rm CO}/dr$ between $R\sim2-16$ kpc, assuming $R_\odot=10$ kpc, from which we obtain $n_{\rm H_2}(R)$ after correcting for $R_\odot$, shown in Figure \ref{fig:nh2}. Interpolating linearly, this gives $n(R_\odot)=0.31\,{\rm cm^{-3}}$. \citet*{sss} also measured CO in the first and second Galactic quadrants within and outside the solar circle. They used the values $R_\odot=10$ kpc and $X=3.6\times 10^{20}$ ${\rm cm^{-2}\, (K^{-1}\, km\, s^{-1})^{-1}}$. Their results for both volume and surface density, corrected to $R_\odot=8.33$ kpc and $X=1.8\times 10^{20}$ ${\rm cm^{-2}\, (K^{-1}\, km\, s^{-1})^{-1}}$, are also shown in Figure \ref{fig:nh2}. In particular, after rescaling and interpolating their volume densities, we have $n(0.95 R_\odot)=0.39\,{\rm cm^{-3}}$. For surface density, we obtain $\Sigma_{\rm H_2+He}=2.7\mppc{2}$. This is the highest value in the literature. \citet{grabelsky87} also measured CO in the outer Galaxy, which, with a 1.8/2.8 correction factor for $X$, as well as correcting $R_\odot$ from 10 to 8.33 kpc, their results near the Sun read $n(1.05R_\odot)= 0.14\,{\rm cm^{-3}}$ and $\Sigma_{\rm H_2}(1.05 R_\odot)= 1.4\mppc{2}$. \citet{digel} also measured $\rm H_2$ in the outer Galaxy. Using his results, we find $n(1.06 R_\odot)=0.13\,{\rm cm^{-3}}$ and $\Sigma_{\rm H_2}(1.06 R_\odot)=2.1\mppc{2}$.

\citet{dame87}, by directly observing clouds within 1 kpc of the Sun only, found local volume density $n_{\rm H_2}=0.10\; {\rm cm^{-3}}$ and surface density $\Sigma_{\rm H_2 + He} = 1.3 \mppc{2}$, which, correcting for $X=2.7$ to $1.8\times 10^{20}{\rm cm^{-2}\, (K^{-1}\, km\, s^{-1})^{-1}}$, gives $0.08\,{\rm cm^{-3}}$ and $0.87 \mppc{2}$. This volume density is lower than many other measurements, and may represent a local fluctuation in the Solar region on a larger scale than the Local Bubble. On the other hand, their surface density value is not the lowest. \citet{luna}, using $X=1.56\times 10^{20}{\rm cm^{-2}\, (K^{-1}\, km\, s^{-1})^{-1}}$, found $\Sigma_{\rm H_2+He}(0.975 R_\odot)=0.24 \mppc{2}$. Correcting for $X$ gives $0.29 \mppc{2}$, which is the lowest value in the literature. However, they admit that their values beyond $0.875\,R_\odot$ are uncertain. Another determination from 2006 \citep{sofue2} gives, after interpolation, $n_{{}_{\rm H_2}}(R_\odot)=0.17\,{\rm cm^{-3}}$ and $\Sigma_{\rm H_2}(R_\odot)=1.4\mppc{2}$, or $\Sigma_{\rm H_2}(R_\odot)=2.0\mppc{2}$.

Figure \ref{fig:nh2} shows the various measurements described here, as well as the overall average and standard error. Although not all measurements are equally certain, in computing average values for $n_{{}_{\rm H_2}}$ and $\Sigma_{\rm H_2}$ we treated all measurements with equal weight. We estimated the resulting uncertainty as the standard deviation divided by the the square root of the number of measurements available at each $R$. We found the mean values and standard errors of volume and surface densities near the Sun to be
\begin{eqnarray}
\label{eq:nh2}
n_{{}_{\rm H_2}}(R_\odot) &=& 0.19 \pm 0.03 \;{\rm cm^{-3}}\\
\Sigma_{\rm H_2+He}(R_\odot) &=& 1.55 \pm 0.32 \mppc{2}.
\end{eqnarray}•
This analysis has not yet taken into account the more recent observations of a significant component of molecular gas that is not associated with CO \citep{heyerdame,hessman}. \citet{planckdg} estimates this ``dark gas'' density to be 118\% that of the CO-associated $\rm H_2$. \citet{herschel2013}, on the other hand, found roughly 40\% at Solar radius. We therefore include the dark molecular gas with a mean value of 79\% and with an uncertainty of 39\%. This gives total molecular gas estimates of
\begin{eqnarray}
\label{eq:sh2}
n_{{}_{\rm H_2+DG}}(R_\odot) &=& 0.34 \pm 0.09 \;{\rm cm^{-3}}\\
\Sigma_{\rm H_2+He+DG}(R_\odot) &=& 2.8 \pm 0.8 \mppc{2}
\end{eqnarray}•which are the values we assume for our analysis. It should be noted, however, that in propagating the errors for dark gas, $n_{{}_{\rm H2}}$ and $\Sigma_{\rm H_2+He}$ always vary together. We take this into account in the statistical analysis by considering only the error on the ratio $\Sigma/\rho$. The same would apply to the error in $X_{\rm CO}$ although this error is much smaller.

Besides the molecular hydrogen's volume density $n_{{}_{\rm H_2}}$ and surface density $\Sigma_{\rm H_2}$, another important quantity is its cloud-cloud velocity dispersion $\sigma_{\rm H_2}$, since this is one of the inputs in the Poisson-Jeans equation. The velocity dispersions of the molecular clouds containing $\rm H_2$ can be inferred from that of their CO, which was found by \citet{lisztburton} to be $\sigma_{\rm H_2}=4.2\pm0.5\;{\rm km\,s^{-1}}$. \citet{belcro} found $\sigma_{\rm CO}=\sigma_{\rm H_2}=3.6\pm0.2\;{\rm km\,s^{-1}}$. \citet{ismss} found $\sigma_{\rm H_2}=3.8\pm2\;{\rm km\,s^{-1}}$. The weighted average of these is approximately given by 

\begin{equation}
\sigma_{\rm H_2}=3.7\pm0.2\;{\rm km\,s^{-1}}.
\end{equation}•

\subsection{The Atomic Hydrogen}
\label{sec:hi}

We now discuss the various measurements of atomic hydrogen HI volume density $n_{{}_{\rm HI}}(z)$ and surface density $\Sigma_{\rm HI}$. These typically are made by observing emissions of hydrogen's 21 cm hyperfine transition. \citet{ismkh} estimate an HI surface density of 8.2 $\mppc{2}$ near the Sun. They separate HI into the Cold Neutral Medium (CNM) and Warm Neutral Medium (WNM).

\begin{figure}[h]
\centering{}\caption{{Atomic hydrogen midplane densities and surface densities determined by various authors between 1978 and 2008.}}
\plotone{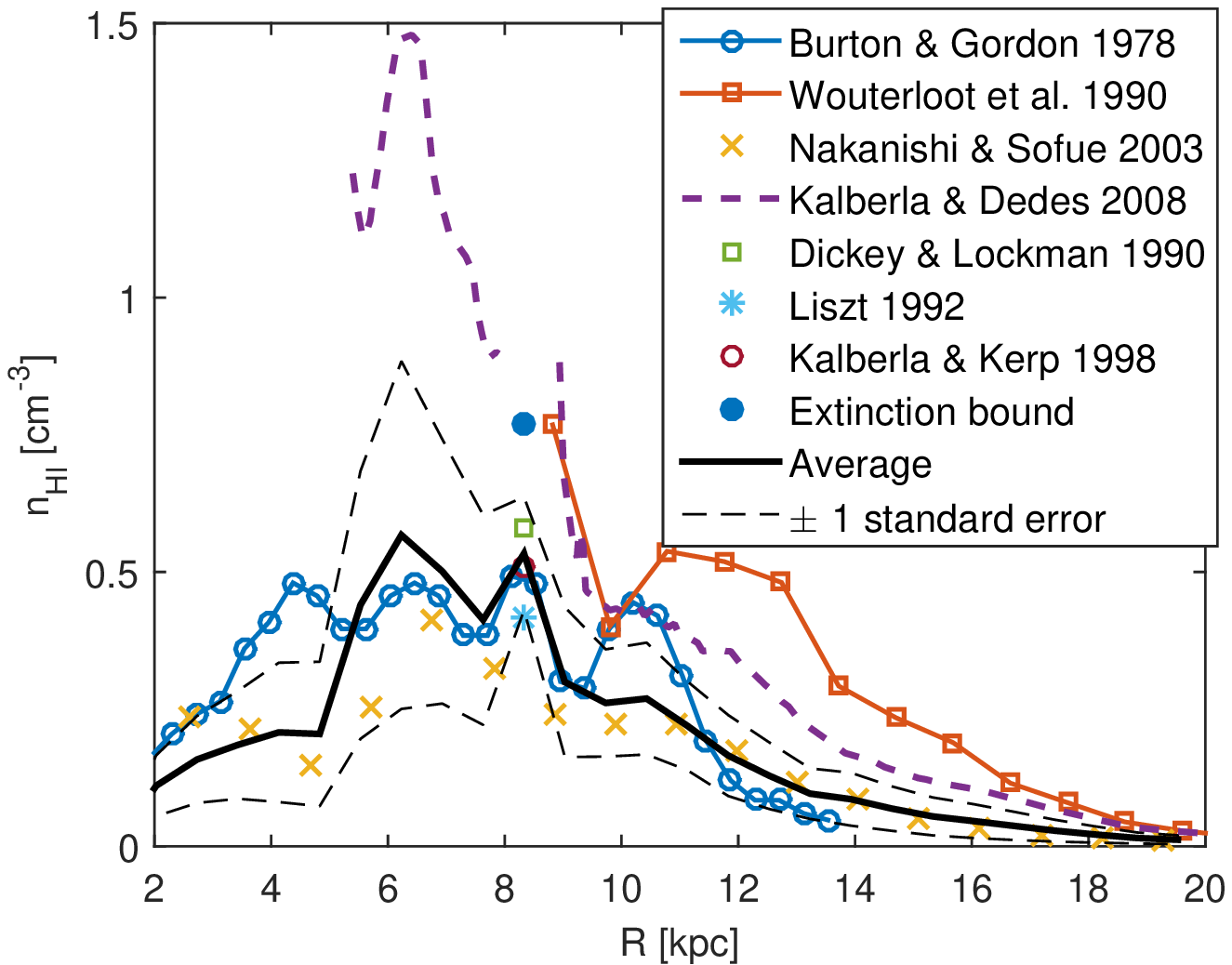}\\
\plotone{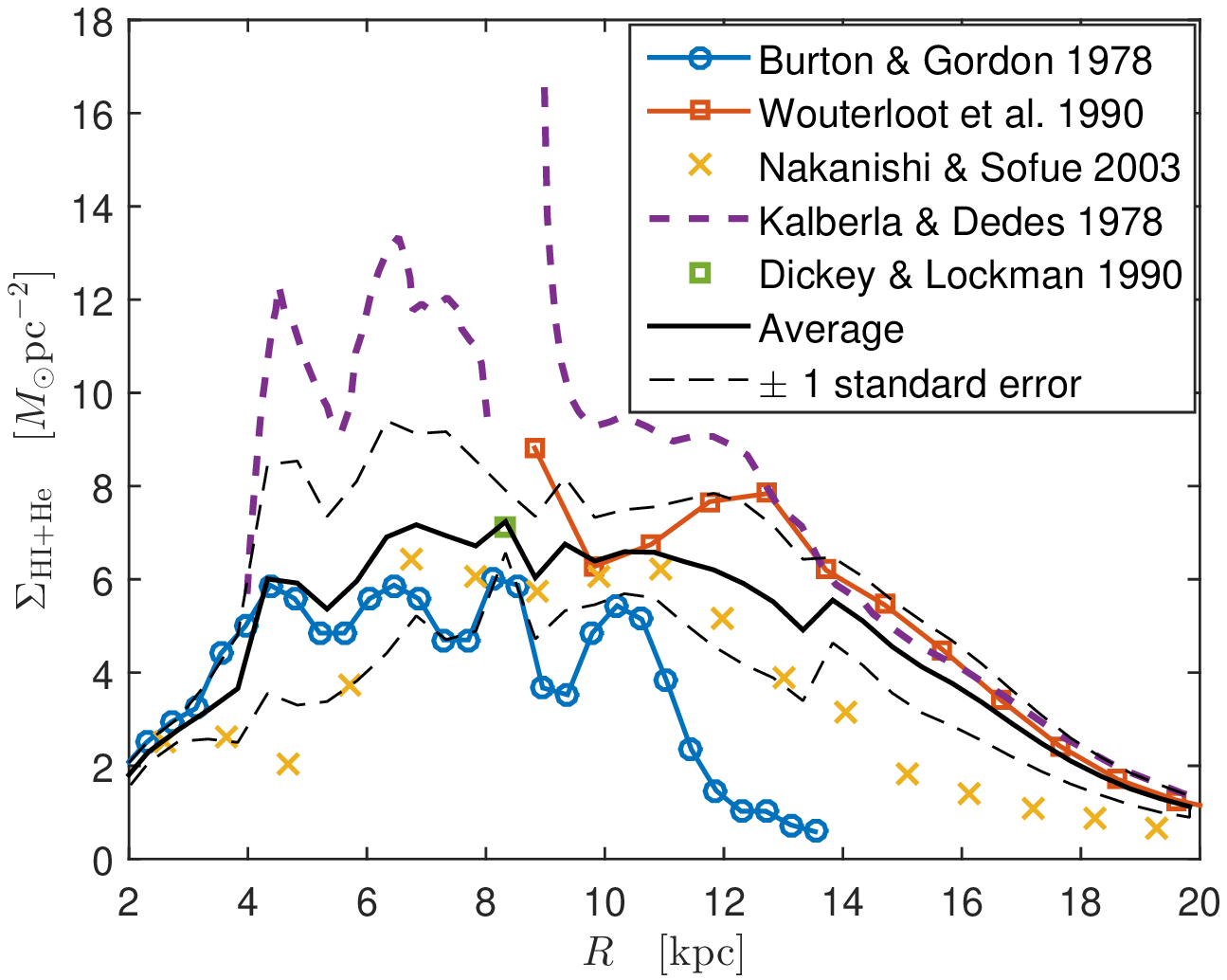}
\label{fig:nhi}
\end{figure}

\citet{dickeylockman}, summarizing several earlier studies, describe the Galactic HI as having approximately constant properties over the range $4\,{\rm kpc}<R<8\,{\rm kpc}$. Their best estimate for the HI parameters over this range is a combination of subcomponents, one thin Gaussian component with  central density $n(0)=0.40\,{\rm cm^{-3}}$ and FWHM = 212 pc (and surface density 2.2$\mppc{2}$), which we identify with the CNM, and a thicker component with central density $n(0)=0.17\,{\rm cm^{-3}}$ and surface density 2.8$\mppc{2}$, which we identify as the WNM. This gives a total of $\Sigma_{\rm HI}=5.0\mppc{2}$, or $\Sigma_{\rm HI+He}=7.1\mppc{2}$. Another measurement is provided by \citet{bg1978}, who measured volume densities for $R\sim 2-16 \,{\rm kpc}$. We interpolate their data (and correct for $R_\odot=10\,{\rm kpc}\to 8.33\,{\rm kpc}$) to obtain $n_{\rm {}_{HI}}=0.49\,{\rm cm^{-3}}$. Although they did not determine surface densities, we can estimate them by assuming a single Gaussian component with FWHM given Dickey \& Lockman (220 - 230 pc). A better estimate is perhaps obtained by assuming, rather than a Gaussian distribution, a distribution with the same shape as Dickey \& Lockman. This amounts to assuming an effective Gaussian FWHM of $\sim 330$ pc. This gives a surface density near the Sun of $\Sigma_{\rm HI+He}=5.9\mppc{2}$. \citet{liszt}, however, argues that the midplane density of Dickey \& Lockman was artificially enhanced to give the correct surface density. He measures midplane density $n_{{}_{\rm HI}}=0.41\,{\rm cm^{-3}}$, which, assuming as for Burton \& Gordon a Gaussian distribution with effective FWHM $330$ pc, gives a surface density of only $\Sigma_{\rm HI+He}=5.1\mppc{2}$. \citet{sofue1} also measured the Galactic HI, from the Galactic center out to $\sim25\,{\rm kpc}$. Their results are shown in Figure \ref{fig:nhi}. Interpolating to $R_\odot$, we have $n_{{}_{\rm HI}}(R_\odot)=0.28\,{\rm cm^{-3}}$ and $\Sigma_{\rm HI + He}=5.9\mppc{2}$, in agreement with the value of Burton \& Gordon.

On the other hand, there are several authors who report much larger mass parameters for Galactic HI. They are \citet{wouterloot} and \citet{kalberladedes}. \citeauthor{wouterloot} used 21 cm observations from outside the Solar circle. Their data are shown in Figure \ref{fig:nhi}. Closest to the Sun, their data show $\Sigma_{{}_{\rm HI+He}}(1.06 R_\odot)=8.6\mppc{2}$ with a FWHM of 300 pc. This corresponds to a midplane density of roughly $n_{{}_{\rm HI}}=0.73\,{\rm cm^{-3}}$. The Kalberla \& Dedes data (also shown in Figure \ref{fig:nhi}) show $\Sigma_{{}_{\rm HI+He}}\simeq10\mppc{2}$. A more refined estimate gives $\Sigma_{{}_{\rm HI+He}}\simeq9\mppc{2}$ \citep{mckee}. This is consistent with a midplane density of roughly $0.8\,{\rm cm^{-3}}$. This is much higher than the value of \citet{kalberlakerp}, who obtained $n_{\rm CNM}=0.3\,{\rm cm^{-3}}$ and $n_{\rm WNM}=0.1\,{\rm cm^{-3}}$. However, there is reason to expect a relatively high HI midplane density. Based on extinction studies, \citet*{bsd1978} find a total hydrogen nucleus density $2n_{{}_{\rm H2}}+n_{{}_{\rm HI}}=1.15 {\rm cm^{-3}}$. Updating this for the newer value of the Galactocentric radius of the Sun $R_\odot$ as in Section \ref{sec:h2}, we have $2n_{{}_{\rm H2}}+n_{{}_{\rm HI}}=1.38\, {\rm cm^{-3}}$. According to the average midplane density determined for molecular hydrogen in Section \ref{sec:h2}, $n_{{}_{\rm H_2}}=0.19\pm0.03\,{\rm cm^{-3}}$, and including an additional $0.15\pm0.07\,{\rm cm^{-3}}$ for the dark molecular hydrogen, we therefore expect an atomic hydrogen density $n_{{}_{\rm HI}}=0.70\pm0.18\,{\rm cm^{-3}}$. Optical thickness corrections, which we explain below, increase this number to 0.84 $\rm cm^{-3}$. The results are shown in Figure \ref{fig:nhi}. As in the case of molecular hydrogen, all measurements were treated with equal weight and the uncertainty was estimated as the standard error at each $R$.

Combining all these results, we have, in the absence of optical thickness corrections,
\begin{eqnarray}
\label{eq:nh21}
n_{{}_{\rm HI}}(R_\odot) &=& 0.53 \pm 0.10 \;{\rm cm^{-3}}\\
\Sigma_{\rm HI+He}(R_\odot) &=& 7.2 \pm 0.7\mppc{2}.
\end{eqnarray}•In the \citet{dickeylockman} model, 70\% of this HI midplane density is in CNM and the remaining 30\% is WNM. In \citet{kalberlakerp}, the numbers are 75\% and 25\%. We will take the average of these two results, 72.5\% and 27.5\%, which give $n_{\rm CNM}=0.38\,{\rm cm^{-3}}$ and $n_{\rm CNM}=0.15\,{\rm cm^{-3}}$

\citet{mckee} pointed out that these numbers must be corrected for the optical depth of the CNM. Assuming the CNM to be optically thin leads to an underestimation of the CNM column density by a factor $\mathcal{R}_{\rm CNM}$. \citet{mckee} estimate this factor to be $\mathcal{R}_{\rm CNM}=1.46$, which they translate, for the total HI column density, to $\mathcal{R}_{\rm HI}=1.20$. Correcting for this gives
\begin{eqnarray}
\label{eq:nh23}
n_{{}_{\rm CNM}} &\simeq& 0.56\, {\rm cm^{-3}}\\
n_{{}_{\rm WNM}} &\simeq& 0.15\, {\rm cm^{-3}}.
\label{eq:nh23a}
\end{eqnarray}•
with totals
\begin{eqnarray}
\label{eq:nh24}
n_{{}_{\rm HI}}(R_\odot) &=& 0.71 \pm 0.13 \;{\rm cm^{-3}}\\
\Sigma_{\rm HI+He}(R_\odot) &=& 8.6 \pm 0.8\mppc{2}.
\label{eq:nh25}
\end{eqnarray}•
which we use for this analysis.

On the other hand, \cite{mckee} argues that the model of \citet{heilesmodel} is more accurate, and recommends increasing the amount of HI in the ISM by a factor of 7.45/6.2.
\citet{mckee}'s values are therefore $n_{\rm CNM}=0.69\,{\rm cm^{-3}}$, $n_{\rm WNM}=0.21\,{\rm cm^{-3}}$, $n_{\rm HI}=0.90\,{\rm cm^{-3}}$, and $\Sigma_{\rm HI}=10.0\pm1.5\mppc{2}$. Although these numbers are different from our average of conventional measurements (Equations \ref{eq:nh23}-\ref{eq:nh25}) , it agrees with the extinction result of \citet{bsd1978} mentioned above once the latter is corrected for the optical depth of the CNM. 
To account for any discrepency, we perform our analysis separately using the values of Equations \ref{eq:nh23} to \ref{eq:nh25} and the results of \citet{mckee}. We present both results in Section \ref{sec:results}.

For the atomic hydrogen's velocity dispersion, \citet{heilestroland}, found $\sigma_{\rm CNM}= 7.1\,{\rm km\,s^{-1}}$ and $\sigma_{\rm WNM}=11.4\,{\rm km\,s^{-1}}$, while \citet{kalberladedes} found $\sigma_{\rm CNM}= 6.1\,{\rm km\,s^{-1}}$ and $\sigma_{\rm WNM}=14.8\,{\rm km\,s^{-1}}$. Earlier, \citet{belcro} measured $\sigma_{\rm HI}=6.9\pm0.4\;{\rm km\,s^{-1}}$, and \citet{dickeylockman} found $\sigma_{\rm HI}=7.0\,{\rm km\,s^{-1}}$ but did not specify if the gas was CNM or WNM. Since these are comparable to more recent measurements of the CNM component of HI, we assume both of these to correspond to $\sigma_{\rm CNM}$. The averages of these values are 
\begin{eqnarray}
\sigma_{\rm CNM}&=& 6.8\pm0.5\,{\rm km\,s^{-1}}\\
\sigma_{\rm WNM}&= &13.1\pm2.4\,{\rm km\,s^{-1}}.
\end{eqnarray}•

\subsection{Ionized Hydrogen}
\label{sec:hii}

Besides the H${}_2$ and the two types of HI (CNM and WNM), there is a fourth, warm, ionized component of interstellar hydrogen, denoted HII. \citet{hf2000} and \citet{hf2006} included this component. \citet{bm} did not include the ionized component in the value for $\Sigma_{\rm ISM}$, possibly because of its very large scale height. Its density is typically obtained by measuring the dispersion of pulsar signals that have passed through the HII clouds. The time delay for a pulse of a given frequency is proportional to the dispersion measure
\begin{equation}
{\rm DM}=\int \! n_e ds
\end{equation}
where the integral is performed along the line of sight to the pulsar, and where $n_e$ is the electron number density, equal to the number density of ionized gas. The dispersion measure perpendicular to the plane of the Galaxy, ${\rm DM}_\perp ={\rm DM}/\sin b$, therefore corresponds to the half surface density $1/2 \;\Sigma_{\rm HII}$. Fitting a spatial distribution (e.g. exponential profile), provides midplane density information. For its midplane density, \citet{ismkh} found $n_{\rm HII}=0.030\;{\rm cm^{-3}}$; \citet{cordes91} found $n_{\rm HII}=0.024\;{\rm cm^{-3}}$; \citet{reynolds91} found $n_{\rm HII}=0.040\;{\rm cm^{-3}}$. The average of these values is
\begin{eqnarray}
\label{eq:hiilarge1}
n_{\rm HII}=0.031\pm0.008\;{\rm cm^{-3}}.
\end{eqnarray}•
This agrees with the traditional model of \citet{taylorcordes}, refined by \citet{cordes}, who found a midplane density of
\begin{eqnarray}
\label{eq:hiilarge2}
n_{\rm HII}=0.034\;{\rm cm^{-3}}.
\end{eqnarray}•
For the HII surface density, \citet{reyn} reports $\Sigma_{\rm HII + He} = 1.57 \mppc{2}$. This is slightly higher that what was found by \citet{taylorcordes}, who found a one-sided column density $1/2 \;N_{\perp,\rm HII}=16.5\,{\rm cm^{-3}\,pc}$, or $\Sigma_{\rm HII+He}=1.1\mppc{2}$, but it is slightly lower than the more recent value of \citet{cordes}, who found $1/2 N_{\perp,\rm HII}=33\,{\rm cm^{-3}\,pc}$, or $\Sigma_{\rm HII + He}=2.3\mppc{2}$. Assuming an exponential profile, with the scale height of 0.9 kpc of \citet{taylorcordes}, the \citet{reyn} result agrees with the midplane densities of Equations \ref{eq:hiilarge1} and \ref{eq:hiilarge2}. However, \citet{gaensler} argued for a scale height of 1.8 kpc that a distribution with midplane density of
\begin{eqnarray}
n_{\rm HII}=0.014\;{\rm cm^{-3}}.
\end{eqnarray}•
Similarly, \citet{schnitzeler} also argues for large scale heights of $\sim 1.4$ kpc. For DM values between 20 and 30 $\rm cm^{-3}\, pc$, this gives a midplane density of $\sim 0.015 \, \rm cm^{-3}\,pc$, as preferred by \citet{mckee}. As we explain in Section \ref{sec:results}, we do not find our model to be consistent with these large scale heights, even without a dark disk. We therefore do not use HII parameters in this paper as a constraint.


For its velocity dispersion, \citet{hf2000} used the value $\sigma_{\rm HII}=40\;{\rm km\,s^{-1}}$. This value seems to have been inferred from scale height measurements of the electrons associated with this ionized gas from \citet{ismkh}. From the data in \citet{reynolds85}, however, we find a turbulent component to the dispersion of only $\sigma_{\rm HII}=21\pm5\;{\rm km\,s^{-1}}$. On the other hand, temperatures between $8,000\,{\rm K}$ and $20,000\;{\rm K}$ give a thermal contribution of $\sigma_{\rm HII, thermal}=\sqrt{2.1\, k_BT/m_p}\simeq12-19\;{\rm km\,s^{-1}}$ \citep[p.14]{ferriere}. Summing these in quadrature gives $\sigma_{\rm HII}=25-29\;{\rm km\,s^{-1}}$. As we will explain in Section \ref{sec:other}, including magnetic and cosmic ray pressure contributions pushes this up to 42 $\rm km\,s^{-1}$. Similarly, \citet{kalb03} also finds $\sigma_{\rm HII}=27\,{\rm km\,s^{-1}}$ while assuming $p_{\rm mag}=p_{\rm cr}=1/3\,p_{\rm turb}$, for a total effective dispersion of $35\,{\rm km\, s^{-1}}$ but did not include a thermal contribution. This gives an average total effective dispersion of $39\pm4\,{\rm km\,s^{-1}}$, which, removing magnetic, cosmic ray, and thermal contributions, gives a turbulent dispersion of $\sigma_{\rm HII}=22\pm3\,{\rm km\,s^{-1}}$. 

The new gas parameter estimates, obtained in this work by incorporating a broad range of literature values, are summarized in Table \ref{tab:oldnew} alongside the old \citep{hf2006} values. The values of \citet{mckee} are also included for comparison.

\begin{table}[h]
\centering{}\caption{{Old values \citep{hf2006} and new values (including all the references mentioned in Section \ref{sec:massmodel}) estimated in this paper. We also include the values of \citet{mckee}.}}
\label{tab:oldnew}
\begin{tabular}{lccc}
\tableline
Component	& \citet{hf2006} 				& This reference	 			& \citet{mckee}\\	
 		& $n(0)$			& $n(0)$			& $n(0)$\\
 		& $[{\rm cm^{-3}}]$	& $[{\rm cm^{-3}}]$	& $[{\rm cm^{-3}}]$\\
\tableline
$\rm H_2{}^*$ 	& 0.30			& $0.19$			& 0.15\\
HI(CNM) 	& 0.46			& 0.56			&0.69\\
HI(WNM) 	& 0.34			& 0.15			&0.21\\
HII		& 0.03			& 0.03			&0.0154\\
\hline
\end{tabular}\\
* does not include dark molecular gas
\end{table}

\subsection{Other Forces}
\label{sec:other}

\citet{bcox} considered the effect of magnetic forces and cosmic ray pressure on the interstellar gas. The effect of the magnetic field is a contribution to the force per unit volume on the $i^{\rm th}$ component of the gas:
\begin{eqnarray}
{\bf f}_{\it i}={\bf J}_{\it i}\times {\bf B}
\end{eqnarray}•
where ${\bf J}_{\it i}$ is the current density associated with gas component $i$, ${\bf B}_i$ is the magnetic induction field due to component $i$, and where
\begin{eqnarray}
{\bf B}\equiv\sum_{i}{\bf B}_{i}
\end{eqnarray}•
is the total magnetic field from all the gas components. Using Ampere's law, we can rewrite the $z$-component of the force as
\begin{eqnarray}
{ f}_{\it z i}&=\frac{1}{\mu_0}\left(\left(\boldsymbol{\nabla}\times{\bf B}_{\it i}\right)\times {\bf B}\right)_z\\
&= \frac{1}{\mu_0} \left({\bf B\cdot div}\right) {\rm  B}_{iz}-\frac{1}{\mu_0}{\bf B}\cdot \frac{\partial{\bf B}_{i}}{\partial z}
\end{eqnarray}•
Since according to \citet{parker}, the magnetic field is, on average, parallel to the plane of the Galaxy, (${\rm B}_z=0$) we will make the approximation that the first term vanishes in equilibrium. The second term couples each gas component to the remaining components, since $\bf B$ represents the total magnetic field. However, summing all components, we have
\begin{eqnarray}
f_z\equiv\sum_i f_{zi}&=-\frac{1}{\mu_0}{\bf B}\cdot \frac{\partial{\bf B}}{\partial z}\\
&= -\frac{\partial}{\partial z}\left(\frac{{\rm B}^2}{2\mu_0}\right).
\end{eqnarray}•
We recognize the form of this expression as the gradient of the magnetic pressure $p_B={\rm B}^2/2\mu_0$. To include this effect in the Poisson-Jeans Equation, we note that the first term on the left-hand-side of Equation \ref{eq:poissonjeans0} has the interpretation (up to an overall mass factor) as the gradient of a `vertical pressure'. This pressure term is a correct description of a population of stars or of gas clouds. In a warm gas, this term has the interpretation as the turbulent pressure of the gas. However, in this case, one also needs to take into account the thermal pressure of the gas
\begin{equation}
p_{\rm thermal}=c_i n_i k_{{}_B}T_i
\end{equation}•
where $c_i$ is a factor that takes into account the degree of ionization of the gas, and $n_i=\rho_i/m_p$ is the number density of the gas atoms. The correct Poisson-Jeans equation in this case therefore reads
\begin{equation}
\frac{\del}{\del z}(\rho_i \sigma_i^2 + \rho_i c_i k_{{}_B} T_i) +  \rho_i \frac{\del \Phi}{\del z} = 0.
\end{equation}•
If we define a `thermal dispersion' as
\begin{equation}
\sigma_{i,T}^2\equiv c_i k_{{}_B} T_i
\end{equation}•
then we can rewrite this as
\begin{equation}
\frac{\del}{\del z}(\rho_i (\sigma_i^2 + \sigma_{i,T}^2)) +  \rho_i \frac{\del \Phi}{\del z} = 0.
\end{equation}•
Clearly, to account for the magnetic pressure, we would include the average of the magnetic pressure term in precisely the same manner:
\begin{equation}
\label{pjmagsum}
\sum_i\frac{\del}{\del z}(\rho_i (\sigma_i^2 + \sigma_{i,T}^2))+ \frac{\del}{\del z}\left\langle\frac{\rm B^2}{2\mu_0}\right\rangle + \rho\frac{\del \Phi}{\del z} = 0.
\end{equation}
where $\rho$ is the total mass density of the gas. In the following subsections, we describe how we model this magnetic pressure term.

\subsubsection{Magnetic Pressure: Thermal Scaling Model}

An important phenomenon noted by \citet{parker} is that the magnetic field B is confined by the weight of the gas through which it penetrates. We therefore would like to solve this equation by following Parker in assuming that the magnetic pressure is proportional to the the thermal pressure term, $p_i=\rho_i c_i k_{{}_B} T_i$. Since each gas component contributes to the total thermal pressure with a different temperature $T_i$, we write:
\begin{eqnarray}
\label{eq:alpha}
\left\langle\frac{{\rm B^2}(z)}{2\mu_0}\right\rangle&=\alpha\sum_i \rho_i(z) \,\sigma_{i,T}^2\\
&= \sum_i \rho_i(z) \,\sigma^2_{i,B}
\end{eqnarray}•
where $\alpha$ is a proportionality constant fixed by $\rm \left\langle B^2(0) \right\rangle$ and $\sum_i\sigma_{i,T}^2$, and where we have defined the `magnetic dispersion' $\sigma^2_{i,B}\equiv\alpha \,\sigma^2_{i,T}$ i.e. the effective dispersion arising from the magnetic pressure. The Poisson-Jeans equation then reads
\begin{equation}
\label{pjmagsum}
\sum_i\frac{\del}{\del z}\bigg(\rho_i \left(\sigma_i^2+\sigma_{i,T}^2+\sigma_{i,B}^2\right)\bigg) + \rho\frac{\del \Phi}{\del z} = 0.
\end{equation}
The above equation admits many solutions. However, we will assume that the unsummed equation
\begin{equation}
\label{pjmag}
\frac{\del}{\del z}\bigg(\rho_i \left(\sigma_i^2+\sigma_{i,T}^2+\sigma_{i,B}^2\right)\bigg) + \rho_i\frac{\del \Phi}{\del z} = 0.
\end{equation}
holds for each component individually. This amounts to assuming that all gas components confine the magnetic field equally. Other solutions can be found by substituting $\sigma_{i,B}^2\to \sigma_{i,B}^2 + S_i(z)$, such that $\sum_i \rho_i(z) S_i(z)=0$. However, if we restrict our analysis to `isothermal' solutions (constant $\sigma_{i,B}^2$) the solution $S_i=0$ will be unique. We can also include the effects of cosmic ray pressure in a similar way, by assuming that the partial cosmic ray pressure is also proportional to the density
\begin{equation}
p_{i,{\rm cr}}(z)=\rho_i(z)\,\sigma_{i,{\rm cr}}^2
\end{equation}
and where $\sigma_{i,{\rm cr}}^2=\beta\, \sigma^2_{i,{\rm T}}$ for some other constant $\beta$. The Poisson-Jeans Equation then reads
\begin{equation}
\frac{\del}{\del z}(\rho_i \sigma_{i,{\rm eff}}^2) = \sum_i\frac{\del}{\del z}\bigg(\rho_i \left(\sigma_i^2+\sigma_{i,T}^2+\sigma_{i,B}^2+\sigma_{i,{\rm cr}}^2\right)\bigg) + \rho\frac{\del \Phi}{\del z} = 0,
\end{equation}
where we have defined
\begin{equation}
\sigma^2_{i,{\rm eff}}=\sigma_i^2+\sigma_{i,T}^2+\sigma_{i,B}^2+\sigma_{i,{\rm cr}}^2.
\end{equation}•
The solution to the Poisson-Jeans Equation for each component will then be
\begin{equation}
\label{eq:ideal1}
\rho_i(z)=\rho_i(0)\,\exp\left(-\frac{\Phi(z)}{\sigma_{i,{\rm eff}}^2}\right).
\end{equation}
Note that since the pressure is additive, the dispersions add in quadrature. \cite{bcox} estimate for the magnetic pressure $p_{B}\simeq (0.4-1.4)\times 10^{-12}\, {\rm dyn}\,{\rm cm^{-2} }$. 
For the cosmic ray pressure, they estimate $p_{\rm cr}\simeq (0.8-1.6)\times 10^{-12}\, {\rm dyn}\,{\rm cm^{-2} }$. 
 The dispersions for each component are shown in Table \ref{tab:disp} for comparison, as well as the effective dispersions in this model.
\begin{table}
\caption{Intrinsic and effective dispersions for ISM components}
\label{tab:disp}
\begin{tabular}{lcccc|cc}
\tableline
\tableline
Component & $\sigma$ & $\sigma_{T}$ & $\sigma_{B}$ & $\sigma_{{\rm cr}}$ &$\sigma_{{\rm eff}}$ (thermal scaling) & $\sigma_{{\rm eff}}$ (warm equipartition)\\
& $[\rm km\, s^{-1}]$ & $[\rm km\, s^{-1}]$ & $[\rm km\, s^{-1}]$ & $[\rm km\, s^{-1}]$ & $[\rm km\, s^{-1}]$ & $[\rm km\, s^{-1}]$\\
\tableline
$\rm H_2    $& 3.7& 0.2& 0.3& 0.3& 3.7& 6.4\\
$\rm HI(CNM)$& 6.8& 0.8& 1.2& 1.3& 7.1&11.8\\
$\rm HI(WNM)$&13.1& 6.7&10.3&10.9&  21&23.7\\
$\rm HII    $&  22&11.8&18.1&  19&36.2&39.9{}
\end{tabular}
\end{table}

There is, however, no clear evidence to support this model. Although $p_{\rm mag}\propto p_{\rm cr} \propto nk_BT$ has been assumed in the past \citep{parker}, this was when the entire gas was treated as a single component. Equation \ref{eq:alpha}, however, is much more specific. We therefore supplement this model with a second model in the next subsection for comparison.

\subsubsection{Magnetic Pressure: Warm Equipartition Model}

Here we describe a second possible model to describe the magnetic and cosmic ray pressures in the interstellar medium. Namely, it has been observed that within the CNM, energy densities in magnetic fields and in turbulence are often roughly equal \citep{heilestroland,heilescrutcher}. Although the ratio between these energies is observed to vary greatly over different molecular clouds, this so-called ``energy equipartition'' seems to be obeyed on average. Physically, this happens because the turbulence amplifies the magnetic field until it becomes strong enough to dissipate through van Alfv\'en waves. Similarly, we expect magnetic fields to trap cosmic rays within the gas until they become too dense and begin to escape. We might therefore expect the cosmic ray and magnetic field energy densities to be similar. For these reasons, an alternative to the first model (Equation \ref{eq:alpha}) would be to assume equipartition of pressure between turbulence, magnetic fields, and cosmic rays:
\begin{equation}
\label{eq:equipartition}
 \sigma_{i}^2=\sigma_{i,{\rm B}}^2=\sigma_{i,{\rm cr}}^2
 \end{equation}
for each component $i$. The effective dispersion, which is the sum of turbulent, magnetic, cosmic ray, and thermal contributions, would then be
\begin{eqnarray}
\sigma_{i,{\rm eff}}^2&=&\sigma_i^2+\sigma_{i,{\rm B}}^2+\sigma_{i,{\rm cr}}^2 + \sigma_{i,T}^2\\
&\simeq&3\,\sigma_{i}^2 + \sigma_{i,T}^2
\label{eq:3s}
\end{eqnarray}•
for each component. One important factor that we should not overlooked here, however, is that the molecular hydrogen and CNM condense to form clouds. Thus, although turbulence, magnetic fields, and cosmic rays may affect the size of the individual clouds, we expect the overall scale height of the cold components to be determined only by the cloud-cloud dispersion and not by these forces. We therefore assume Equations \ref{eq:equipartition}-\ref{eq:3s} only for the warm components WNM and HII. For the cold components ($\rm H_2$ and CNM), we assume $\sigma_{i,{\rm eff}}=\sigma_{i}$. We perform calculations separately for the two different magnetic field and cosmic ray models. The effective dispersions in both models are shown in Table \ref{tab:disp}. As we will see, the results from both models are in good agreement with one another.

\section{Results and Discussion}
\label{sec:results}
We now present the results of the analysis described above.
Using the midplane densities of Section \ref{sec:massmodel}, we calculate according to the Poisson-Jeans equation the corresponding $\rm H_2$ and HI 
surface densities, and from these we compute the chi-squared value, $(\Sigma-\hat{\Sigma})^2/\Delta_{\Sigma}^2$, from the disagreement between these values and the measured values. This is done over a range of values for $\Sigma_D$ and $h_D$. We thereby determine the regions of parameter space where the disagreement exceeds the 68\% and 95\% bounds, as will be displayed in the plots below. The scale height $h_D$ is defined such that
\begin{equation}
\rho(Z\!\!=\!h_D)=\rho(Z\!\!=\!0)\,{\rm sech^2}(1/2).
\end{equation}• 

\begin{figure}[t]
\caption{Confidence bounds on DDDM parameter space as a function of $h_D$, the dark disk ${\rm sech}^2(z/2h_D)$ scale height, using averages and uncertainties from Sections \ref{sec:h2} to \ref{sec:hii}. Solid lines represent 95\% bounds and dashed lines represent 68\% bounds.}
\label{fig:nomag_kramer}
\plotone{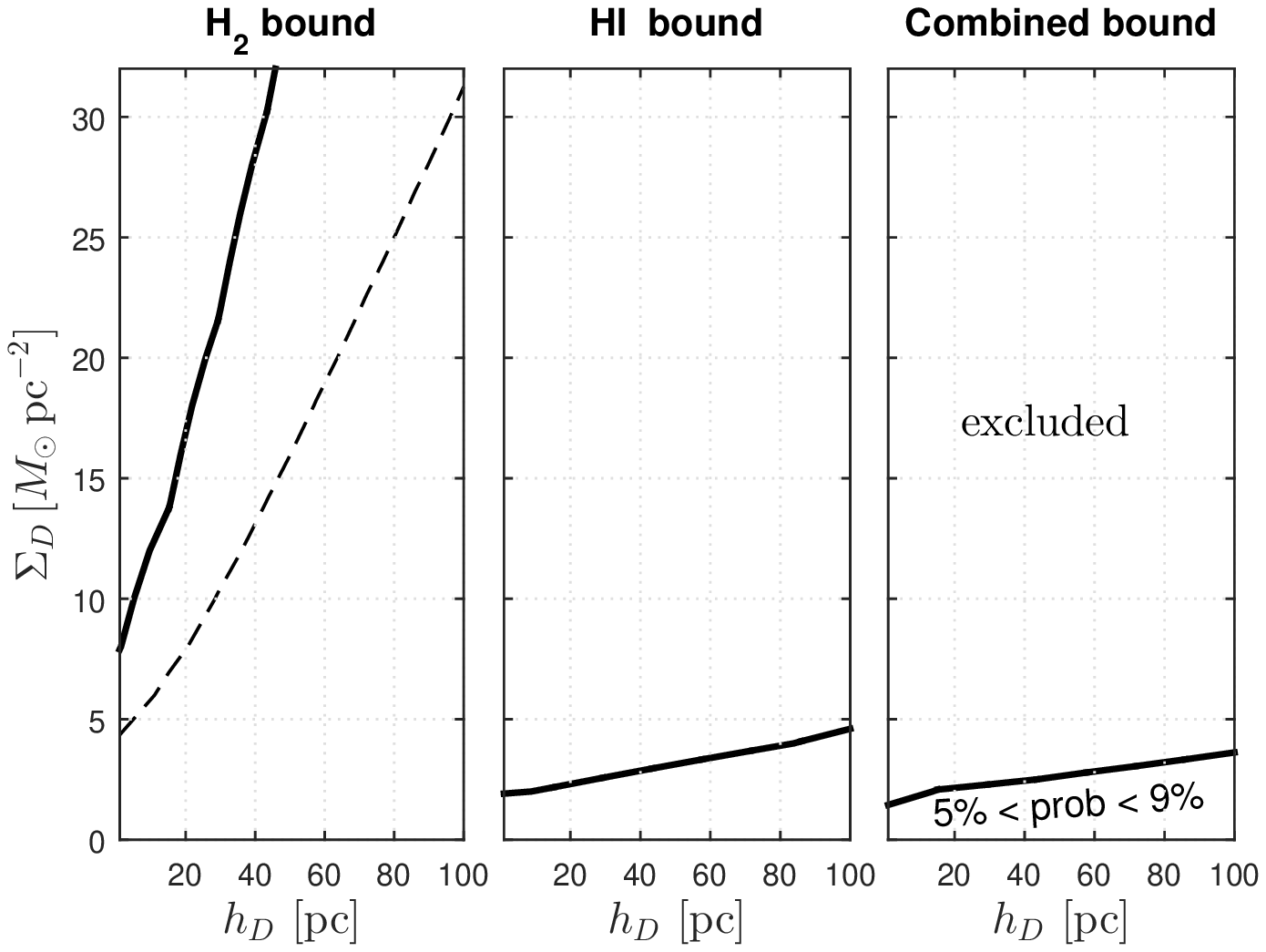}
\end{figure}
We begin by determining the bounds without including the contribution from magnetic fields and cosmic rays. The result is shown in Figure \ref{fig:nomag_kramer}. The uncertainty $\rm H_2$ is dominated by that of the dark molecular gas, while the uncertainty of HI is dominated by that of the WNM velocity dispersion (18\%). We can see that although the $\rm H_2$ parameters are consistent with dark disk surface densities of (for low scale height) up to $10-12\mppc{2}$, the HI parameters point toward lower surface densities, and that the combined probabilities are lower than 9\% for all models. These results make it apparent that the model without magnetic fields is inconsistent.
\begin{figure}[h]
\centering{}\caption{Bounds on DDDM parameter space as in Figure \ref{fig:nomag_kramer}, but including contributions from magnetic fields and from cosmic rays. $Black$: computed assuming `thermal scaling model'. $Red$: computed assuming `warm equipartition model'.}
\label{fig:mag_kramer}
\plotone{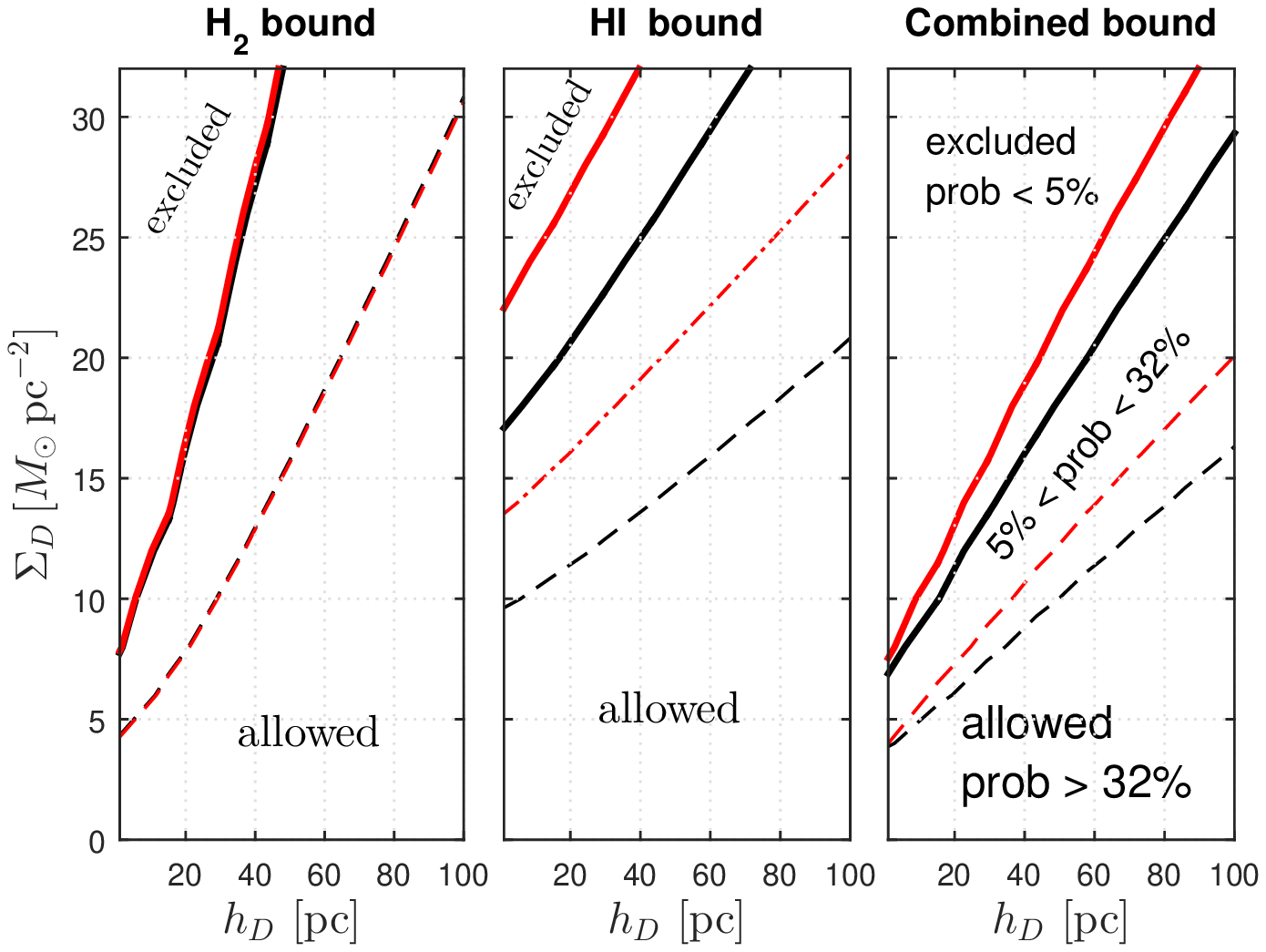}
\end{figure}
%

On the other hand, when we include the pressure contribution from the magnetic fields and from cosmic rays, we find that both the HI and $\rm H_2$ parameters allow non-zero surface densities $\Sigma_D$, with an upper bound of 
$\Sigma_D\simeq10\mppc{2}$ in both models, for low scale heights. Higher scale heights are consistent with even higher dark disk surface densities. The results are shown in Figure \ref{fig:mag_kramer}.

\begin{figure}[t]
\caption{Confidence bounds as in Figure \ref{fig:nomag_kramer} but using the values of \citet{mckee}. $Left$: Not including magnetic field and cosmic ray contributions. $Right$: including magnetic field and cosmic ray contributions as in Figure \ref{fig:mag_kramer}.}
\label{fig:mckee}
\plottwo{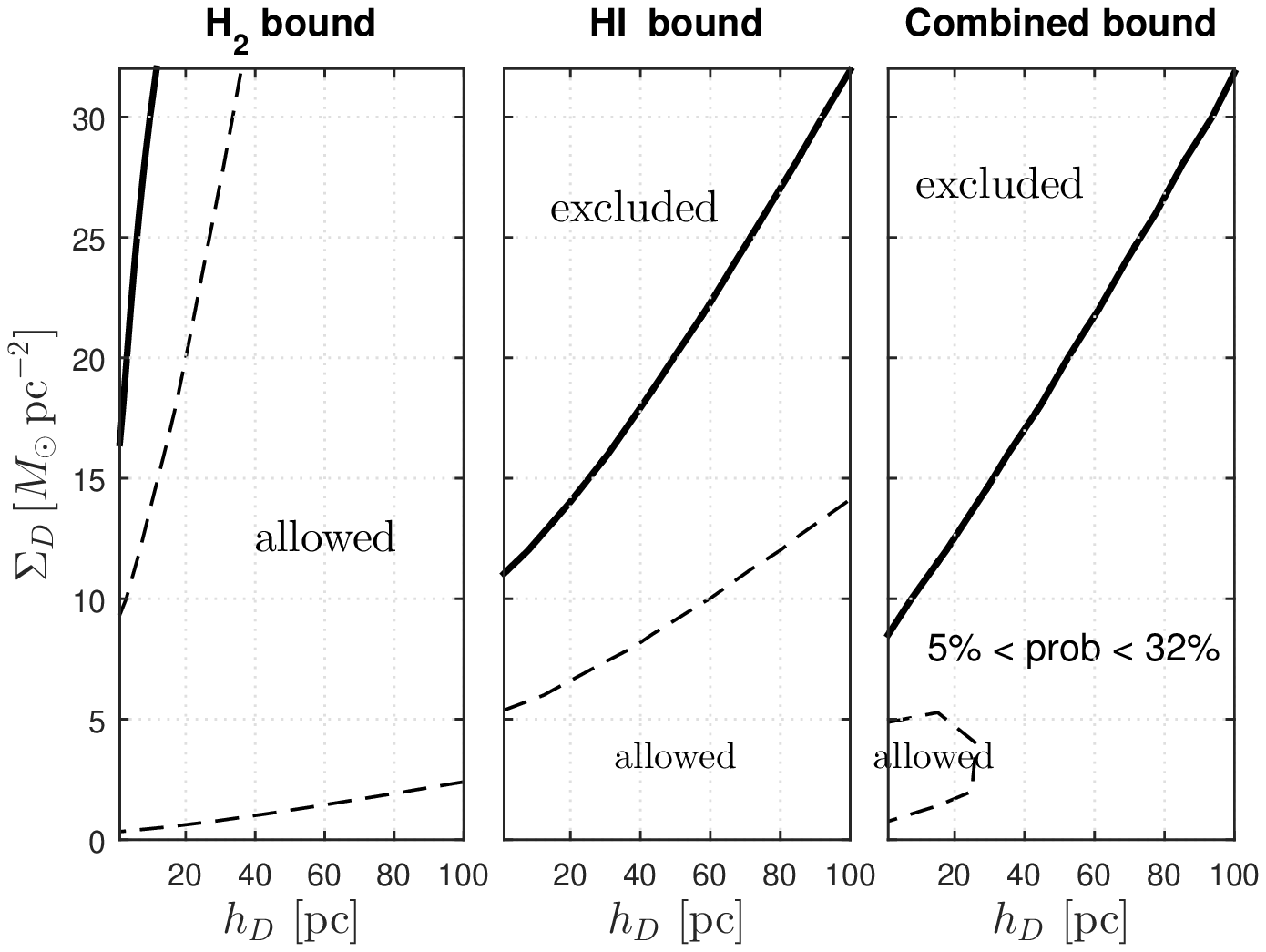}{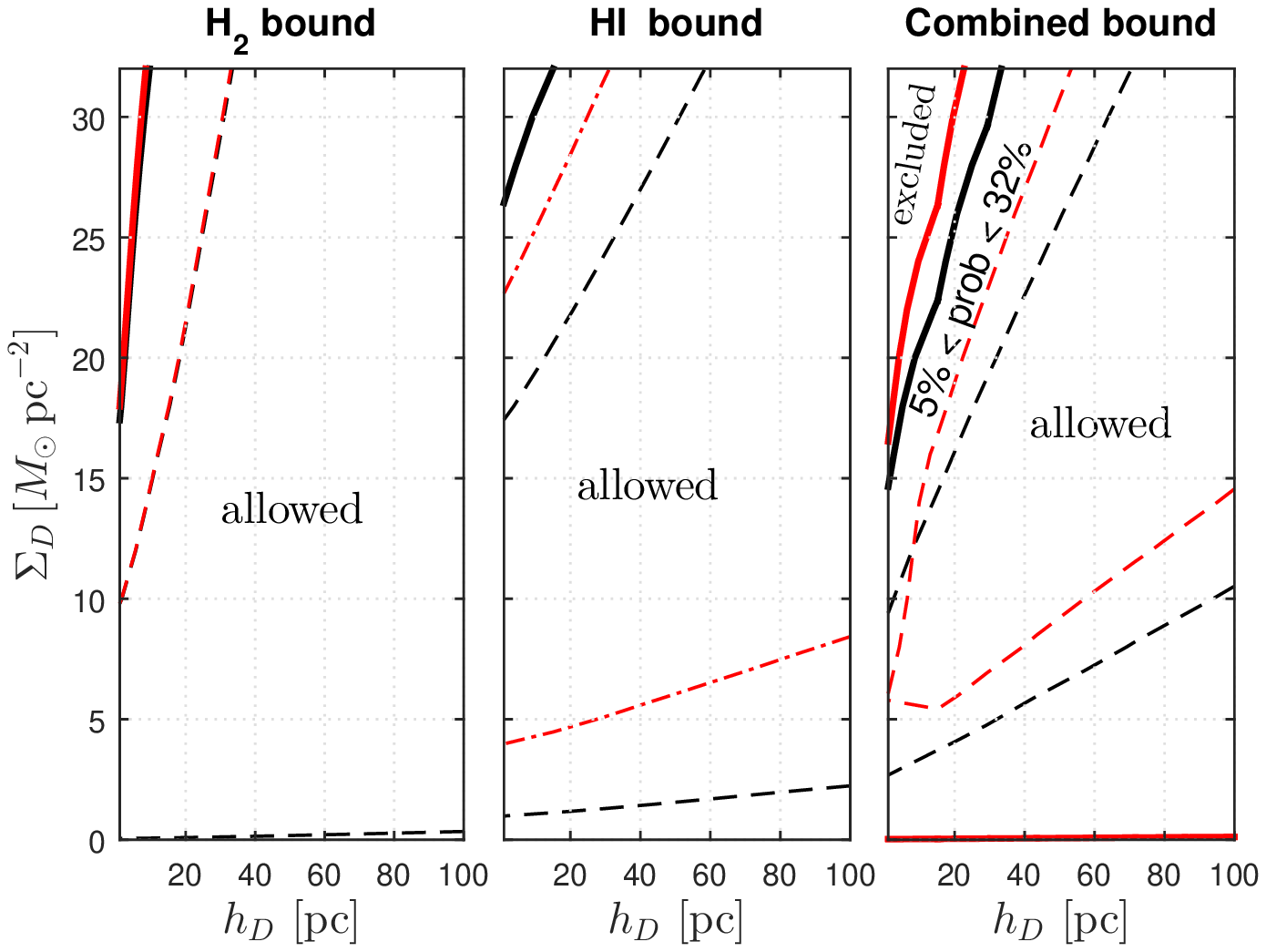}
\end{figure}

For comparison, we also include the corresponding results using the values of \citet{mckee}. Using these values and including magnetic fields and cosmic ray contributions, the data favors a non-zero surface density for the dark disk of between 5 and 15$\mppc{2}$. Note that when neglecting magnetic field and cosmic rays pressures, only low dark disk surface densities seem consistent with the data.

\subsubsection*{Ionized Hydrogen Results and Issues}

As was mentioned in Section \ref{sec:hii}, various authors have measured DM values for the HII component of the Milky Way in the range 20-30 $\rm cm^{-3}\,pc$. Older models favored low scale height with midplane densities as high as $0.034\,{\rm cm^{-3}}$, while newer models favor large scale height models with midplane densities as low as $0.014\,{\rm cm^{-3}}$. However, using our model, the results of our Poisson-Jeans solver are consistent with only low scale heights. Following the models described in Section \ref{sec:other}, we find scale heights for the HII of 0.9 kpc for the thermal scaling model and 1.0 kpc for the warm equipartition model, assuming $\Sigma_D=0$. Incorporating a more massive dark disk makes these scale heights smaller. Possible reasons for this might be:\\
1) The magnetic field model must be modified to include a different value of $\alpha$ for HII. This could be in correspondence with the result of \citet{beuermann}, who found that Galactic magnetic fields contained two components, one with short scale height and one with larger scale height. These two components would likely be described by different $\alpha$. We could then attribute the low scale height component to the molecular and atomic gas and the large scale height component to the ionized gas. We know of no such alternative in the warm equipartition model.\\
2) The isothermal assumption may not be valid for HII.  In fact, as explained in \citet{gaensler}, the volume filling fraction of HII may also vary a lot with scale height. If this is the case, then it would be incorrect to treat the HII as an isothermal component as the degrees of freedom that the temperature describes (the gas clouds) vary with distance from the Galactic midplane.

\subsubsection*{Stability Issues and Kinematic Constraints}

In Figure 7 we show the bound we obtained from the kinematics of A stars in the Solar region, accounting for nonequilibrium features of the population, namely a net displacement and vertical velocity relative to the Galactic midplane. We also note that there will exist disk stability bounds. The true analysis is subtle, but a step toward the analysis is done by \citet{shaviv16_2} who develops the stability criterion for a heterogeneous Milky Way disk including a thin dark matter disk. We convert his bound to a bound in the $h_D-\Sigma_D$ plane and superimpose this bound on the gas parameter bound of the present work. We see that a disk with significant mass ($\Sigma_D$) and $h_D > 30$ pc is consistent with all current bounds.

In addition to stability issues, \citet{hessman} has argued that there exist other issues with using the vertical Jeans equation to constrain the dynamical mass in the MW disk. In particular, spiral structure must be taken into account when performing these analyses. Indeed, \citet{shaviv16_1} has pointed out that the effect of spiral arm crossing is to induce a `ringing' in the dynamics of tracer stars. However, the present analysis assumes that the time scales for this ringing are much shorter in gas components so that the analysis is valid. Spiral arm crossing could also induce non-equilibrium features in the tracer population, such as discussed in \citet{paper1}, but as in \citeauthor{paper1}, including this effect would allow for more dark matter.
%

\begin{figure}[t]
\caption{The red shaded region, delimited by the solid red line, denotes the parameters allowed by the stability bound of \citet{shaviv16_2}. The blue shaded region, delimited by the solid blue line, denotes the parameters allowed by the kinematic bound of \citet{paper1}. The grey shaded region, delimited by the solid black line, denotes the parameters allowed by the gas parameters as determined in the current paper. As in Figure \ref{fig:mckee}, the dashed and solid black lines denote the 68\% and 95\% bounds obtained from the combined gas bound, including magnetic field and cosmic ray contributions, and using the parameters of \citet{mckee}.}
\label{fig:stability}
\plotone{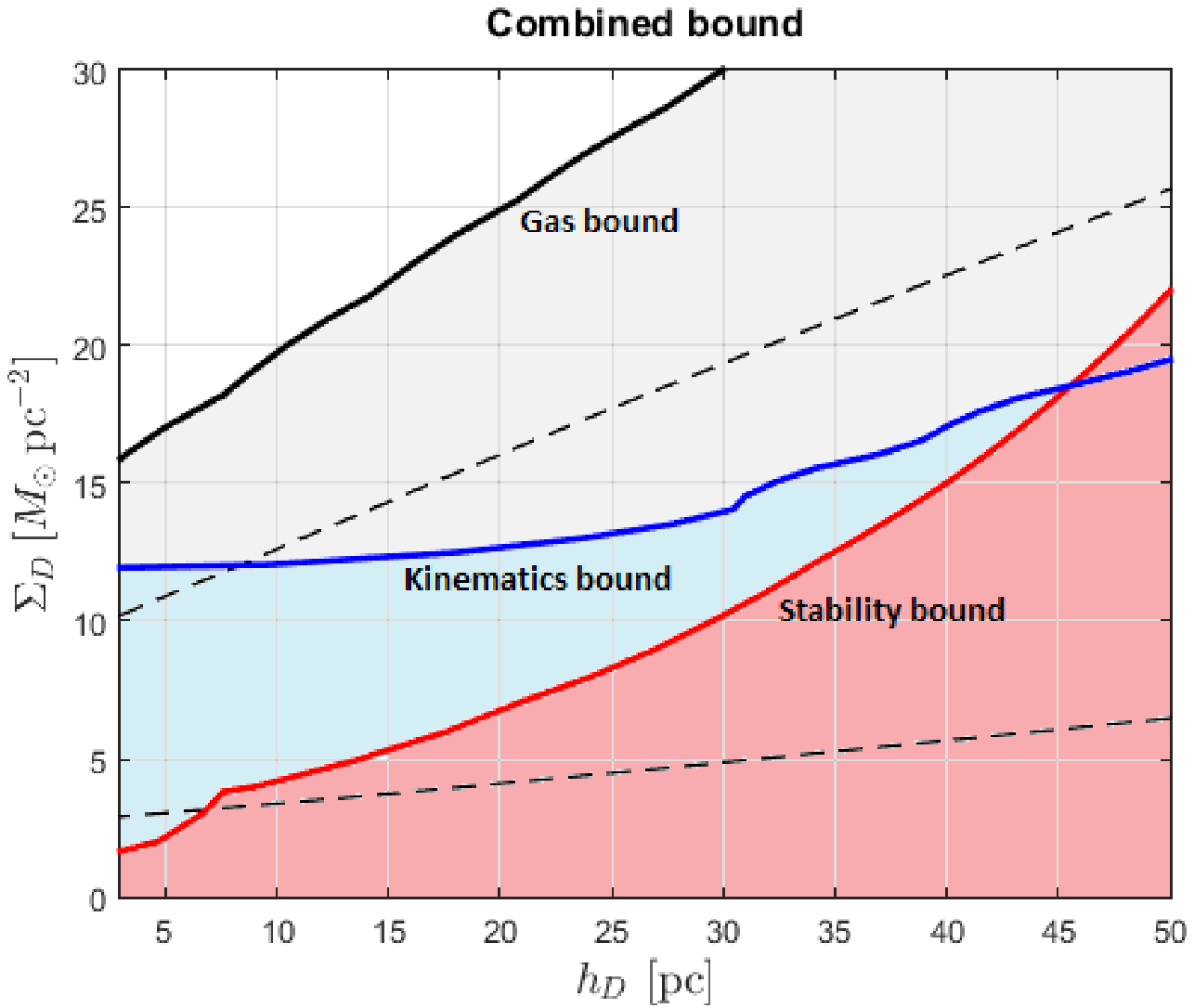}
\end{figure}

\section{Conclusion}
In this paper we have shown how to use measured midplane and surface densities of various galactic plane components to constrain or discover a dark disk. Although literature values of atomic hydrogen midplane densities are discordant, their mean value is consistent with the remaining gas parameters when magnetic and cosmic ray pressures are included. Using the global averages of literature values of gas parameters that we compiled, we find the data are consistent with dark disk surface densities as high as $10\mppc{2}$ for low scale height, and as low as zero. The gas parameters of \citet{mckee} seem to favor an even higher non-zero dark disk surface density. Current data are clearly inadequate to decide this definitively. Further measurements of visible and dark $\rm H_2$ density and WNM density and dispersion, as well as further refinements of magnetic field and cosmic ray models for cold gas could allow placing more robust bounds on a dark disk.

\acknowledgements

We would like to thank Chris Flynn and Chris McKee for their comments and suggestions, and also for reviewing our results. We would also like to thank Johann Holmberg and Jo Bovy for useful discussions. Thanks also to Doug Finkbeiner, Jo Bovy, Alexander Tielens, Katia Ferri\`ere, and Matt Walker for help on the ISM parameters. We would also like to thank our reviewer for their detailed review of our work. EDK was supported by NSF grants of LR and by Harvard FAS, Harvard Department of Physics, and Center for the Fundamental Laws of Nature. LR was supported by NSF grants PHY-0855591 and PHY-1216270. Calculations performed in MATLAB 2015a.


\begin{thebibliography}{}
\expandafter\ifx\csname natexlab\endcsname\relax\def\natexlab#1{#1}\fi

\bibitem[{{Bahcall}(1984{\natexlab{a}})}]{bahc84c}
{Bahcall}, J.~N. 1984{\natexlab{a}}, \apj, 287, 926

\bibitem[{{Bahcall}(1984{\natexlab{b}})}]{bahc84b}
---. 1984{\natexlab{b}}, \apj, 276, 169

\bibitem[{{Bahcall}(1984{\natexlab{c}})}]{bahc84a}
---. 1984{\natexlab{c}}, \apj, 276, 156

\bibitem[{{Belfort} \& {Crovisier}(1984)}]{belcro}
{Belfort}, P., \& {Crovisier}, J. 1984, \aap, 136, 368

\bibitem[{{Beuermann} {et~al.}(1985){Beuermann}, {Kanbach}, \&
  {Berkhuijsen}}]{beuermann}
{Beuermann}, K., {Kanbach}, G., \& {Berkhuijsen}, E.~M. 1985, \aap, 153, 17

\bibitem[{{Binney} \& {Merrifield}(1998)}]{bm}
{Binney}, J., \& {Merrifield}, M. 1998, {Galactic Astronomy}

\bibitem[{{Bohlin} {et~al.}(1978){Bohlin}, {Savage}, \& {Drake}}]{bsd1978}
{Bohlin}, R.~C., {Savage}, B.~D., \& {Drake}, J.~F. 1978, \apj, 224, 132

\bibitem[{{Boulares} \& {Cox}(1990)}]{bcox}
{Boulares}, A., \& {Cox}, D.~P. 1990, \apj, 365, 544

\bibitem[{{Bronfman} {et~al.}(1988){Bronfman}, {Cohen}, {Alvarez}, {May}, \&
  {Thaddeus}}]{bronfman}
{Bronfman}, L., {Cohen}, R.~S., {Alvarez}, H., {May}, J., \& {Thaddeus}, P.
  1988, \apj, 324, 248

\bibitem[{{Burton} \& {Gordon}(1978)}]{bg1978}
{Burton}, W.~B., \& {Gordon}, M.~A. 1978, \aap, 63, 7

\bibitem[{{Clemens} {et~al.}(1988){Clemens}, {Sanders}, \& {Scoville}}]{css}
{Clemens}, D.~P., {Sanders}, D.~B., \& {Scoville}, N.~Z. 1988, \apj, 327, 139

\bibitem[{{Cordes} \& {Lazio}(2002)}]{cordes}
{Cordes}, J.~M., \& {Lazio}, T.~J.~W. 2002, ArXiv Astrophysics e-prints,
  astro-ph/0207156

\bibitem[{{Cordes} {et~al.}(1991){Cordes}, {Weisberg}, {Frail}, {Spangler}, \&
  {Ryan}}]{cordes91}
{Cordes}, J.~M., {Weisberg}, J.~M., {Frail}, D.~A., {Spangler}, S.~R., \&
  {Ryan}, M. 1991, \nat, 354, 121

\bibitem[{{Dame} {et~al.}(2001){Dame}, {Hartmann}, \& {Thaddeus}}]{dht}
{Dame}, T.~M., {Hartmann}, D., \& {Thaddeus}, P. 2001, \apj, 547, 792

\bibitem[{{Dame} {et~al.}(1987){Dame}, {Ungerechts}, {Cohen}, {de Geus},
  {Grenier}, {May}, {Murphy}, {Nyman}, \& {Thaddeus}}]{dame87}
{Dame}, T.~M., {Ungerechts}, H., {Cohen}, R.~S., {et~al.} 1987, \apj, 322, 706

\bibitem[{{Dickey} \& {Lockman}(1990)}]{dickeylockman}
{Dickey}, J.~M., \& {Lockman}, F.~J. 1990, \araa, 28, 215

\bibitem[{{Digel}(1991)}]{digel}
{Digel}, S.~W. 1991, PhD thesis, Harvard University, Cambridge, MA.

\bibitem[{{Draine}(2011)}]{draine}
{Draine}, B.~T. 2011, {Physics of the Interstellar and Intergalactic Medium}

\bibitem[{{Fan} {et~al.}(2013){Fan}, {Katz}, {Randall}, \& {Reece}}]{dddm}
{Fan}, J., {Katz}, A., {Randall}, L., \& {Reece}, M. 2013, Physics of the Dark
  Universe, 2, 139

\bibitem[{{Ferri{\`e}re}(2001)}]{ferriere}
{Ferri{\`e}re}, K.~M. 2001, Reviews of Modern Physics, 73, 1031

\bibitem[{{Flynn} {et~al.}(2006){Flynn}, {Holmberg}, {Portinari}, {Fuchs}, \&
  {Jahrei{\ss}}}]{hf2006}
{Flynn}, C., {Holmberg}, J., {Portinari}, L., {Fuchs}, B., \& {Jahrei{\ss}}, H.
  2006, \mnras, 372, 1149

\bibitem[{{Gaensler} {et~al.}(2008){Gaensler}, {Madsen}, {Chatterjee}, \&
  {Mao}}]{gaensler}
{Gaensler}, B.~M., {Madsen}, G.~J., {Chatterjee}, S., \& {Mao}, S.~A. 2008,
  \pasa, 25, 184

\bibitem[{{Gillessen} {et~al.}(2009){Gillessen}, {Eisenhauer}, {Trippe},
  {Alexander}, {Genzel}, {Martins}, \& {Ott}}]{gillesen}
{Gillessen}, S., {Eisenhauer}, F., {Trippe}, S., {et~al.} 2009, \apj, 692, 1075

\bibitem[{{Grabelsky} {et~al.}(1987){Grabelsky}, {Cohen}, {Bronfman},
  {Thaddeus}, \& {May}}]{grabelsky87}
{Grabelsky}, D.~A., {Cohen}, R.~S., {Bronfman}, L., {Thaddeus}, P., \& {May},
  J. 1987, \apj, 315, 122

\bibitem[{{Heiles} \& {Crutcher}(2005)}]{heilescrutcher}
{Heiles}, C., \& {Crutcher}, R. 2005, in Lecture Notes in Physics, Berlin
  Springer Verlag, Vol. 664, Cosmic Magnetic Fields, ed. R.~{Wielebinski} \&
  R.~{Beck}, 137

\bibitem[{{Heiles} {et~al.}(1981){Heiles}, {Kulkarni}, \&
  {Stark}}]{heilesmodel}
{Heiles}, C., {Kulkarni}, S., \& {Stark}, A.~A. 1981, \apjl, 247, L73

\bibitem[{{Heiles} \& {Troland}(2003)}]{heilestroland}
{Heiles}, C., \& {Troland}, T.~H. 2003, \apj, 586, 1067

\bibitem[{{Hessman}(2015)}]{hessman}
{Hessman}, F.~V. 2015, \aap, 579, A123

\bibitem[{{Heyer} \& {Dame}(2015)}]{heyerdame}
{Heyer}, M., \& {Dame}, T.~M. 2015, \araa, 53, 583

\bibitem[{{Holmberg} \& {Flynn}(2000)}]{hf2000}
{Holmberg}, J., \& {Flynn}, C. 2000, \mnras, 313, 209

\bibitem[{{Kalberla}(2003)}]{kalb03}
{Kalberla}, P.~M.~W. 2003, \apj, 588, 805

\bibitem[{{Kalberla} \& {Dedes}(2008)}]{kalberladedes}
{Kalberla}, P.~M.~W., \& {Dedes}, L. 2008, \aap, 487, 951

\bibitem[{{Kalberla} {et~al.}(2007){Kalberla}, {Dedes}, {Kerp}, \&
  {Haud}}]{kalb07}
{Kalberla}, P.~M.~W., {Dedes}, L., {Kerp}, J., \& {Haud}, U. 2007, \aap, 469,
  511

\bibitem[{{Kalberla} \& {Kerp}(1998)}]{kalberlakerp}
{Kalberla}, P.~M.~W., \& {Kerp}, J. 1998, \aap, 339, 745

\bibitem[{{Kramer} \& {Randall}(2016)}]{paper1}
{Kramer}, E.~D., \& {Randall}, L. 2016, \apj, 824, 116

\bibitem[{{Kulkarni} \& {Heiles}(1987)}]{ismkh}
{Kulkarni}, S.~R., \& {Heiles}, C. 1987, in Astrophysics and Space Science
  Library, Vol. 134, Interstellar Processes, ed. D.~J. {Hollenbach} \& H.~A.
  {Thronson}, Jr., 87--122

\bibitem[{{Liszt}(1992)}]{liszt}
{Liszt}, H.~S. 1992, in Astrophysics and Space Science Library, Vol. 180, The
  Center, Bulge, and Disk of the Milky Way, ed. L.~{Blitz}, 111--130

\bibitem[{{Liszt} \& {Burton}(1983)}]{lisztburton}
{Liszt}, H.~S., \& {Burton}, W.~B. 1983, in Astrophysics and Space Science
  Library, Vol. 100, Kinematics, Dynamics and Structure of the Milky Way, ed.
  W.~L.~H. {Shuter}, 135--142

\bibitem[{{Luna} {et~al.}(2006){Luna}, {Bronfman}, {Carrasco}, \& {May}}]{luna}
{Luna}, A., {Bronfman}, L., {Carrasco}, L., \& {May}, J. 2006, \apj, 641, 938

\bibitem[{{McKee} {et~al.}(2015){McKee}, {Parravano}, \& {Hollenbach}}]{mckee}
{McKee}, C., {Parravano}, A., \& {Hollenbach}, D.~J. 2015, \apj

\bibitem[{{Nakanishi} \& {Sofue}(2003)}]{sofue1}
{Nakanishi}, H., \& {Sofue}, Y. 2003, Publ. Astron. Soc. Jap., 55, 191

\bibitem[{{Nakanishi} \& {Sofue}(2006)}]{sofue2}
---. 2006, Publ. Astron. Soc. Jap., 58, 847

\bibitem[{{Okumura} {et~al.}(2009){Okumura}, {Kamae}, \& {for the Fermi LAT
  Collaboration}}]{okumura09}
{Okumura}, A., {Kamae}, T., \& {for the Fermi LAT Collaboration}. 2009, ArXiv
  e-prints, arXiv:0912.3860

\bibitem[{{Oort}(1932)}]{oort1}
{Oort}, J.~H. 1932, \bain, 6, 249, 6, 249

\bibitem[{{Oort}(1960)}]{oort2}
---. 1960, \bain, 15, 45, 15, 45

\bibitem[{{Parker}(1966)}]{parker}
{Parker}, E.~N. 1966, \apj, 145, 811

\bibitem[{{Pineda} {et~al.}(2013){Pineda}, {Langer}, {Velusamy}, \&
  {Goldsmith}}]{herschel2013}
{Pineda}, J.~L., {Langer}, W.~D., {Velusamy}, T., \& {Goldsmith}, P.~F. 2013,
  \aap, 554, A103

\bibitem[{{Planck Collaboration} {et~al.}(2011){Planck Collaboration}, {Ade},
  {Aghanim}, {Arnaud}, {Ashdown}, {Aumont}, {Baccigalupi}, {Balbi}, {Banday},
  {Barreiro}, \& et~al.}]{planckdg}
{Planck Collaboration}, {Ade}, P.~A.~R., {Aghanim}, N., {et~al.} 2011, \aap,
  536, A19

\bibitem[{{Randall} \& {Reece}(2014)}]{dino}
{Randall}, L., \& {Reece}, M. 2014, \prl, 112, 161301

\bibitem[{{Read}(2014)}]{readrev}
{Read}, J.~I. 2014, Journal of Physics G Nuclear Physics, 41, 063101

\bibitem[{{Read} {et~al.}(2008){Read}, {Lake}, {Agertz}, \& {Debattista}}]{dd1}
{Read}, J.~I., {Lake}, G., {Agertz}, O., \& {Debattista}, V.~P. 2008, \mnras,
  389, 1041

\bibitem[{{Reynolds}(1985)}]{reynolds85}
{Reynolds}, R.~J. 1985, \apj, 294, 256

\bibitem[{{Reynolds}(1991)}]{reynolds91}
{Reynolds}, R.~J. 1991, in IAU Symposium, Vol. 144, The Interstellar Disk-Halo
  Connection in Galaxies, ed. H.~{Bloemen}, 67--76

\bibitem[{{Reynolds}(1992)}]{reyn}
{Reynolds}, R.~J. 1992, in American Institute of Physics Conference Series,
  Vol. 278, American Institute of Physics Conference Series, 156--165

\bibitem[{{Sanders} {et~al.}(1984){Sanders}, {Solomon}, \& {Scoville}}]{sss}
{Sanders}, D.~B., {Solomon}, P.~M., \& {Scoville}, N.~Z. 1984, \apj, 276, 182

\bibitem[{{Schnitzeler}(2012)}]{schnitzeler}
{Schnitzeler}, D.~H.~F.~M. 2012, \mnras, 427, 664

\bibitem[{{Scoville} \& {Sanders}(1987)}]{ismss}
{Scoville}, N.~Z., \& {Sanders}, D.~B. 1987, in Astrophysics and Space Science
  Library, Vol. 134, Interstellar Processes, ed. D.~J. {Hollenbach} \& H.~A.
  {Thronson}, Jr., 21--50

\bibitem[{{Shaviv}(2016{\natexlab{a}})}]{shaviv16_1}
{Shaviv}, N.~J. 2016{\natexlab{a}}, ArXiv e-prints, arXiv:1606.02595

\bibitem[{{Shaviv}(2016{\natexlab{b}})}]{shaviv16_2}
---. 2016{\natexlab{b}}, ArXiv e-prints, arXiv:1606.02851

\bibitem[{{Taylor} \& {Cordes}(1993)}]{taylorcordes}
{Taylor}, J.~H., \& {Cordes}, J.~M. 1993, \apj, 411, 674

\bibitem[{{Wouterloot} {et~al.}(1990){Wouterloot}, {Brand}, {Burton}, \&
  {Kwee}}]{wouterloot}
{Wouterloot}, J.~G.~A., {Brand}, J., {Burton}, W.~B., \& {Kwee}, K.~K. 1990,
  \aap, 230, 21

\end{thebibliography}

\end{document}